\documentclass[12pt,onecolumn]{IEEEtran}
\IEEEoverridecommandlockouts

\usepackage[dvips]{graphicx}
\usepackage{psfrag}

\usepackage[cmex10]{amsmath}
\usepackage{amssymb}
\usepackage{amsthm}
\usepackage{amsfonts}

\usepackage{cases}
\usepackage{units}
\usepackage{dsfont}

\usepackage{cite}
\usepackage{url}

\newcommand{\E}{\mathrm{E}}

\newcommand{\VAR}{\mathrm{VAR}}
\newcommand{\st}{\;|\;}

\providecommand{\norm}[1]{\lVert{#1}\rVert}

\DeclareMathOperator{\dB}{dB}
\DeclareMathOperator{\Prob}{Pr}

\DeclareMathOperator{\HB}{HB}

\DeclareMathOperator{\NoSI}{No-SI}

\DeclareMathOperator{\WZ}{WZ}

\newtheorem{prop}{Proposition}
\newtheorem{lemma}{Lemma}
\newtheorem{corr}{Corollary}

\begin{document}

\title{Minimum Expected Distortion in Gaussian Source Coding with Fading Side Information}

\author{Chris~T.~K.~Ng,~\IEEEmembership{Member,~IEEE,}
	Chao~Tian,~\IEEEmembership{Member,~IEEE,}\\
        Andrea~J.~Goldsmith,~\IEEEmembership{Fellow,~IEEE,}
        and~Shlomo~Shamai~(Shitz),~\IEEEmembership{Fellow,~IEEE}
\thanks{This work was supported by the US Army under MURI award W911NF-05-1-0246,
the ONR under award N00014-05-1-0168, DARPA's ITMANET program under grant 1105741-1-TFIND,
and a grant from Intel.
The work of C.~Ng was supported by a Croucher Foundation Fellowship.
The work of S.~Shamai has been supported  by the European Commission in the framework of the FP7 Network of Excellence in Wireless COMmunications NEWCOM++.
The material in this paper was presented in part at the IEEE Information Theory Workshop, Lake Tahoe, CA, September 2007.}%
\thanks{C.~Ng was with the Department of Electrical Engineering, Stanford University, Stanford, CA 94305 USA (e-mail: Chris.Ng@ieee.org).}%
\thanks{C.~Tian is with AT\&T Labs--Research, Florham Park, NJ 07932 USA (e-mail: tian@research.att.com).}%
\thanks{A.~Goldsmith is with the Department of Electrical Engineering, Stanford University, Stanford, CA 94305 USA (e-mail: andrea@wsl.stanford.edu).}%
\thanks{S.~Shamai~(Shitz) is with the Department of Electrical Engineering, Technion - Israel Institute of Technology, Haifa 32000, Israel (e-mail: sshlomo@ee.technion.ac.il).}%
}

\maketitle
\thispagestyle{empty}

\begin{abstract}

An encoder, subject to a rate constraint, wishes to describe a Gaussian source under squared error distortion. The decoder, besides receiving the encoder's description, also observes side information consisting of uncompressed source symbol subject to slow fading and noise. The decoder knows the fading realization but the encoder knows only its distribution. The rate--distortion function that simultaneously satisfies the distortion constraints for all fading states was derived by Heegard and Berger. A layered encoding strategy is considered in which each codeword layer targets a given fading state. When the side-information channel has two discrete fading states, the expected distortion is minimized by optimally allocating the encoding rate between the two codeword layers. For multiple fading states, the minimum expected distortion is formulated as the solution of a convex optimization problem with linearly many variables and constraints. Through a limiting process on the primal and dual solutions, it is shown that single-layer rate allocation is optimal when the fading probability density function is continuous and quasiconcave (e.g., Rayleigh, Rician, Nakagami, and log-normal). In particular, under Rayleigh fading, the optimal single codeword layer targets the least favorable state as if the side information was absent.

\end{abstract}


\section{Introduction}
\label{sec:intro}

\IEEEPARstart{I}{n} lossy data compression, side information at the decoder can help reduce the distortion in the reconstruction of the source \cite{wyner76:rate_dist_side_info}.
The decoder, however, may have access to the side information only through an unreliable channel.
For example, in distributed compression over wireless sensor networks, correlated sensor measurements from a neighboring node may be available to the decoder through a fading wireless channel.
In this work, we consider a Gaussian source where the encoder is subject to a rate constraint and the distortion metric is the mean squared error of the reconstruction.
In addition to the compressed symbol, we assume that the decoder observes the original symbol through a separate analog fading channel.
We assume, similar to the approach in \cite{shamai03:bc_app_slow_fade_mimo}, that the fading is quasistatic, and that the decoder knows the fading realization but the encoder knows only its distribution.
The rate--distortion function that dictates the rate required to satisfy the distortion constraint associated with each fading state is given by Heegard and Berger in \cite{heegard85:rate_dist_si}.
We consider a layered encoding strategy based on the uncertain fading realization in the side-information channel, and optimize the rate allocation among the possible fading states to minimize the expected distortion.
In particular, we formulate the distortion minimization as a convex optimization problem, and develop an efficient representation for the Heegard--Berger rate--distortion function under which the optimization problem size is linear in the number of discrete fading states.
Furthermore, we identify the conditions under which single-layer rate allocation is expected-distortion--minimizing,
and extend these optimality conditions for continuous fading distributions through a limiting process on the primal and dual solutions in the optimization.
We show that single-layer rate allocation is optimal for fading distributions with continuous, quasiconcave probability density functions such as Rayleigh, Rician, Nakagami, and log-normal.

When the side-information channel exhibits no fading, the distortion is given by the Wyner--Ziv rate--distortion function \cite{wyner78:rate_dist_si_gen}.
Rate--distortion is considered in \cite{gray73:bounds_cond_RD, zamir96:rate_loss_WZ} when the side information is also available at the encoder, and in \cite{fleming06:RD_mixed_type_info} when there is a combination of decoder-only and encoder-and-decoder side information.
Successive refinement source coding in the presence of side information is considered in \cite{steinberg04:succ_refine_WZ, tian07:ms_suc_ref_deg_si}.
The side-information scalable rate--distortion region is characterized in \cite{tian08:si_scal_src_cod}, in which the user with inferior side information decodes an additional layer of the source-coding codeword.
Lossless source coding with an unknown amount of side information at the decoder is considered in \cite{feder02:src_bc_unkn_rx_si}, in which a fixed data block is broadcast to different users in a variable number of channel uses \cite{shulman00:static_broadcast}.
In \cite{tian08:sr_bc_exp_dist_gaus, ng09:dist_min_glbc_sr}, expected distortion is minimized in the transmission of a Gaussian source over a slowly fading channel in the absence of channel state information at the transmitter (CSIT).
Broadcast transmission with imperfect CSIT is considered in \cite{steiner08:imperf_tx_si_bc}.
Another application of source coding with uncertain side information is in systematic lossy source-channel coding \cite{shamai98:sys_lossy_src_ch} over a fading channel without CSIT.\@
For example, when upgrading legacy communication systems, a digital channel may be added to augment an existing analog channel.
In this case the analog reception then plays the role of side information in the decoding of the description from the digital channel.
In \cite{gunduz08:wz_bc_hda, nayak10:wz_bc_digital}, hybrid digital/analog and digital transmission schemes are considered for Wyner--Ziv coding over broadcast channels.
The system model studied in this paper is also related to distributed source coding over multiple links \cite{ishwar05:rate_constrn_dist_est, chen08:rob_dist_src_cod} where,
besides source coding over a finite-capacity reliable link, noisy versions of the source are described through additional backhaul links with infinite capacity but that are subject to random failure.
At the decoder, the realized quality of the side information is determined by the number of backhaul links that are successfully connected.
Similar models are considered in \cite{simeone11:unreliable_backhaul_links} for distributed unreliable relay communications.

The remainder of the paper is organized as follows.
The system model is described in Section~\ref{sec:sysmod}.
Section~\ref{sec:min_exp_dist} derives the minimum expected distortion and presents the convex optimization framework when the side-information channel has discrete fading states.
Section~\ref{sec:rate_alloc_fading} investigates the optimal rate allocation under different fading distributions in the side-information channel.
Section~\ref{sec:single_layer_rate_alloc} considers the optimality of single-layer rate allocation under discrete fading states as well as continuous fading distributions.
Conclusions are given in Section~\ref{sec:conclu}.

\section{System Model}
\label{sec:sysmod}

\subsection{Source Coding with Fading Side-Information Channel}

\begin{figure}
  \centering
  \psfrag{RX}[][]{$R_X$}
  \psfrag{X}[r][r]{$\{X\}$}
  \psfrag{Xh}[l][l]{$\{\hat{X}\}$}
  \psfrag{E[D]}[l][l]{$\E[D]$}
  \psfrag{S=F(s)}[r][r]{$S\sim F(s)$}
  \psfrag{Y=rSX+Z}[l][l]{$Y'=\sqrt{S}X+Z$}
  \psfrag{Z=N(0,1)}[l][l]{$Z\sim\mathcal{N}(0,1)$}
  \psfrag{X=N(0,sx)}[l][l]{$X\sim\mathcal{N}(0,\sigma_X^2)$}
  \includegraphics{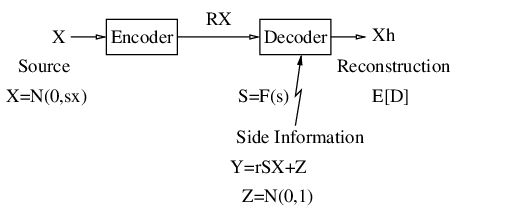}
  \caption{Source coding with fading side-information channel.}
  \label{fig:src_coding_si}
\end{figure}

Consider the system model shown in Fig.~\ref{fig:src_coding_si}.
An encoder wishes to describe a real Gaussian source sequence $\{X\}$ under a rate constraint of $R_X$ nats per symbol, where the sequence of random variables are independent identically distributed (i.i.d.) with $X \sim \mathcal{N}(0,\sigma_X^{2})$.
The decoder, in addition to receiving the encoder's description, observes side information $Y'$, where $Y'=\sqrt{S}X+Z$, with $Z\sim$ i.i.d. $\mathcal{N}(0,1)$.
Hence the quality of the side information depends on $S$, the power gain of the side-information channel.
We assume $S$ is a quasistatic random variable:
it is drawn from some cumulative distribution function (cdf) $F(s)$ at the beginning of each transmission block and remains unchanged through the block.
The decoder knows the realization of $S$, but the encoder knows only its distribution $F(s)$.
When the fading distribution is continuous, it is characterized by the probability density function (pdf) $f(s)=F'(s)$.
The decoder forms an estimate of the source and reconstructs the sequence $\{\hat{X}\}$.
We are interested in minimizing the expected squared error distortion $\E[D]$ of the reconstruction, where $D=(X-\hat{X})^2$.

Suppose the side-information channel has $M$ discrete fading states.
Let the probability distribution of $S$ be given as follows:
\begin{align}
\label{eq:pmf_s_p}
\Pr\{S = s_i\} = p_i,\quad i=1,\dotsc,M, \qquad \sum_{i=1}^M p_i = 1
\end{align}
where the $s_i$'s are enumerated in ascending order $0 \leq s_1 < s_2 < \dotsb < s_M$.
Let $Y'_i$ denote the side information under fading state $s_i$
\begin{align}
Y'_i \triangleq \sqrt{s_i}X + Z,\qquad i=1,\dotsc,M.
\end{align}
Note that the set of side information random variables are stochastically degraded.
Let $\hat{X}_i$ be the reconstruction when side information $Y'_i$ is available at the decoder, and $D_i$ be the corresponding squared error distortion.
The minimum expected distortion under rate constraint $R_X$ is then given by
\begin{align}
\label{eq:E_D_opt_RD}
\E[D]^* = \min_{\mathbf{D}\,:\: R(\mathbf{D})\leq R_X} \mathbf{p}^T\mathbf{D}
\end{align}
where
$\mathbf{p} \triangleq [p_1 \dots p_M]^T$,
$\mathbf{D} \triangleq [D_1 \dots D_M]^T$,
and $R(\mathbf{D})$ is the rate--distortion function that simultaneously satisfies the distortion set $\mathbf{D}$.

\subsection{Heegard--Berger Rate--Distortion Function}

The rate--distortion function that dictates the rate required to simultaneously satisfy a set of distortion constraints associated with a set of degraded side-information random variables is given by Heegard and Berger in \cite{heegard85:rate_dist_si} (an alternate form for $M=2$ is described in \cite{kaspi94:RD_SI_may_present}).
When the side information random variables satisfy the degradedness condition
$X \leftrightarrow Y_M \leftrightarrow Y_{M-1} \leftrightarrow \dotsb \leftrightarrow Y_1$,
the rate--distortion function is
\begin{align}
\label{eq:R_HB_D}
R_{\HB}(\mathbf{D}) = \min_{W_1^M \in P(\mathbf{D})}\, \sum_{i=1}^{M} I(X;W_i|Y_i,W_1^{i-1})
\end{align}
where $W_1^i$ denotes the vector $W_1,\dotsc,W_i$.
The minimization takes place over $P(\mathbf{D})$, the set of all $W_1^M$ jointly distributed with $X, Y_1^M$ such that:
\begin{align}
W_1^M \leftrightarrow X \leftrightarrow Y_M \leftrightarrow Y_{M-1} \leftrightarrow \dotsb \leftrightarrow Y_1
\end{align}
and there exists decoding functions $\hat{X}_i(Y_i, W_1^i)$'s under given distortion measures $d_i$'s that satisfy
\begin{align}
\label{eq:E_d_Xi_Xhi_Di}
\E[d_i(X,\hat{X}_i)] \leq D_i,\qquad i=1,\dotsc,M.
\end{align}

As noted in \cite{heegard85:rate_dist_si}, since $R_{\HB}(\mathbf{D})$ depends on $X,Y_1^M$ only through the marginal distribution $p(x,y_i)$, $i=1,\dotsc,M$, the degradedness of the side information need not be physical.
We construct $Y_1^M$ to have the same marginals as ${Y'}_1^M$ by setting $p(y_i|x) = p(y'_i|x)$, $i=1,\dotsc,M$.
The rate--distortion function $R(\mathbf{D})$ in (\ref{eq:E_D_opt_RD}) is then given by the Heegard--Berger rate--distortion function (\ref{eq:R_HB_D}) with squared error distortion measures $d_i(X,\hat{X}_i) = (X-\hat{X}_i)^2$.

\section{Minimum Expected Distortion}
\label{sec:min_exp_dist}

\subsection{Gaussian Source under Squared Error Distortion}

The Heegard--Berger rate--distortion function $R_{\HB}(D_1,\dotsc,D_M)$ for a Gaussian source under squared error distortion is given in \cite{heegard85:rate_dist_si, tian07:ms_suc_ref_deg_si,  tian08:si_scal_src_cod}.
For $M=2$, \cite{heegard85:rate_dist_si} describes the Gaussian rate--distortion function where the worst fading state corresponds to no side information, and \cite{tian07:ms_suc_ref_deg_si} considers side information with different quality levels.
The Gaussian Heegard--Berger rate--distortion function is considered in \cite{tian08:si_scal_src_cod} for $M>2$.
However, the representations of $R_{\HB}(D_1,\dotsc,D_M)$ described in \cite{heegard85:rate_dist_si, tian07:ms_suc_ref_deg_si} are characterized by exponentially-many distinct regions, and \cite{tian08:si_scal_src_cod} involves optimal Gaussian random variables whose variances are determined by an algorithmic procedure.
These characterizations, though complete, are not amenable to efficient minimization of the expected distortion.
In this section, we derive a representation for $R_{\HB}(D_1,\dotsc,D_M)$ that can be incorporated in an optimization framework.
In particular, instead of describing the achievable distortion set by its exponentially-many segments of boundaries,
we construct a characterization that comprises a sequence of convex inequalities, each relating the achievable distortion between two adjacent fading states.
Consequently, we formulate the distortion minimization as a convex optimization problem where the number of variables and constraints are linear in $M$.
First we consider the case when the side-information channel has only two discrete fading states ($M=2$); in Section~\ref{sec:mult_disc_fading} we extend the analysis to multiple fading states where $M>2$.

When the side-information channel has two fading states, the encoder constructs a source-coding scheme that consists of two layers of codewords.
The base layer is designed to be decodable under either channel condition, while the top layer is only decodable under the more favorable channel realization.
We derive the rate requirements of the two codeword layers, and optimally allocate the encoding rate $R_X$ between them to minimize the expected distortion.
For $M=2$, the Heegard--Berger rate--distortion function is given by
\begin{align}
\label{eq:R_HB_D1D2}
R_{\HB}(D_1,D_2) &= \min_{W_1,W_2 \in P(D_1,D_2)}\, \{I(X;W_1|Y_1) + I(X;W_2|Y_2,W_1)\}.
\end{align}
For a Gaussian source under a squared error distortion measure, a jointly Gaussian codebook is optimal \cite{heegard85:rate_dist_si, tian07:ms_suc_ref_deg_si, tian08:si_scal_src_cod}.
When $W_1^M,X$ are jointly Gaussian, the mutual information expressions in (\ref{eq:R_HB_D1D2}) evaluate to
\begin{align}
&I(X;W_1|Y_1) + I(X;W_2|Y_2,W_1)\notag\\
&= \frac{1}{2}\log(\VAR[X|Y_1])-\frac{1}{2}\log\frac{\VAR[X|Y_1,W_1]}{\VAR[X|Y_2,W_1]}-\frac{1}{2}\log(\VAR[X|Y_2,W_1,W_2])\\
\label{eq:log_VAR_ds_XYW}
&= -\frac{1}{2}\log(s_1+\sigma_x^{-2})-\frac{1}{2}\log\bigl(1+(s_2-s_1)\VAR[X|Y_1,W_1]\bigr)-\frac{1}{2}\log(\VAR[X|Y_2,W_1,W_2])
\end{align}
where $\log$ is the natural logarithm, and (\ref{eq:log_VAR_ds_XYW}) follows from expanding the conditional variance expressions by applying Lemma~\ref{lem:VAR_XYW_XW_s} and Corollary~\ref{corr:Y_sX_Z_ij}.
The proof of Lemma~\ref{lem:VAR_XYW_XW_s} is given in Appendix~\ref{sec:proof_VAR_XYW_XW_s}.

\begin{lemma}
\label{lem:VAR_XYW_XW_s}
Let $X,W_1^k$ be jointly Gaussian random variables.
If $Y=\sqrt{s}X+Z$, where $Z\sim\mathcal{N}(0,1)$ is independent from $X,W_1^k$, then
\begin{align}
\VAR[X|Y,W_1^k] = \bigl(\VAR[X|W_1^k]^{-1}+s\bigr)^{-1}.
\end{align}
\end{lemma}

\begin{corr}
\label{corr:Y_sX_Z_ij}
Let $Y_j = \sqrt{s_j}X+Z$, $Y_i = \sqrt{s_i}X+Z$.
\begin{align}
\frac{\VAR[X|Y_i,W_1^k]}{\VAR[X|Y_j,W_1^k]} = 1 + (s_j-s_i)\VAR[X|Y_i,W_1^k].
\end{align}
\end{corr}

To characterize the Heegard--Berger rate--distortion function $R_{\HB}(D_1,D_2)$, we substitute (\ref{eq:log_VAR_ds_XYW}) in (\ref{eq:R_HB_D1D2}), and minimize over $W_1, W_2$
\begin{align}
\label{eq:R_HB_W1_W2}
\begin{split}
R_{\HB}(D_1,D_2) &= -\frac{1}{2}\log(s_1+\sigma_x^{-2})
+\min_{W_1} \Bigl\{ -\frac{1}{2}\log\bigl(1+(s_2-s_1)\VAR[X|Y_1,W_1]\bigr)\\
&\qquad +\min_{W_2} \bigl\{ -\frac{1}{2}\log(\VAR[X|Y_2,W_1,W_2])\bigr\} \Bigr\}.
\end{split}
\end{align}
Note that $s_2>s_1\geq0$ by assumption.
Accordingly, in the inner minimization in (\ref{eq:R_HB_W1_W2}), $R_{\HB}(D_1,D_2)$ is decreasing in $\VAR[X|Y_2,W_1,W_2]$.
Hence the choice of $W_2$ is optimal when $\VAR[X|Y_2,W_1,W_2]$ is increased until one of its upper bound constraints is tight
\begin{align}
\label{eq:W2_VAR_D2}
\max_{W_2}\, \VAR[X|Y_2,W_1,W_2] = \min(\VAR[X|Y_2,W_1],\, D_2).
\end{align}
The optimal $W_2^*$ that achieves (\ref{eq:W2_VAR_D2}) is presented subsequently.
The first term in the $\min(\cdot)$ expression in (\ref{eq:W2_VAR_D2}) follows from the non-negativity of the mutual information $I(X;W_2|Y_2,W_1)$, and the second one follows from the distortion constraint on $\hat{X}_2$ as given in (\ref{eq:E_d_Xi_Xhi_Di})
\begin{align}
\VAR[X|Y_2,W_1,W_2] = \E\bigl[\bigl(X-\hat{X}_2(Y_2,W_1,W_2)\bigr)^2\bigr] \leq D_2.
\end{align}
Applying Corollary~\ref{corr:Y_sX_Z_ij}, the first term in (\ref{eq:W2_VAR_D2}) evaluates to
\begin{align}
\label{eq:VAR_XY2W1_XY1W1_s}
\VAR[X|Y_2,W_1] &= \bigl(\VAR[X|Y_1,W_1]^{-1}+s_2-s_1\bigr)^{-1}.
\end{align}
Under optimal $W_2$, therefore, the Heegard--Berger rate--distortion function in (\ref{eq:R_HB_W1_W2}) reduces to
\begin{align}
\label{eq:R_HB_W1}
\begin{split}
R_{\HB}(D_1,D_2) &= -\frac{1}{2}\log(s_1+\sigma_x^{-2})
+\min_{W_1} \Bigl\{ -\frac{1}{2}\log\bigl(1+(s_2-s_1)\VAR[X|Y_1,W_1]\bigr)\\
&\qquad -\frac{1}{2}\log \min\Bigl(\bigl(\VAR[X|Y_1,W_1]^{-1}+s_2-s_1\bigr)^{-1},\, D_2\Bigl)
\Bigr\}.
\end{split}
\end{align}
The maximization over $W_1$ in (\ref{eq:R_HB_W1}) has a similar structure as the one previously considered in (\ref{eq:W2_VAR_D2}).
Specifically, $R_{\HB}(D_1,D_2)$ in (\ref{eq:R_HB_W1}) is decreasing in $\VAR[X|Y_1,W_1]$.
Hence $W_1$ is optimal when $\VAR[X|Y_1,W_1]$ is increased until it meets one of its upper bound constraints
\begin{align}
\label{eq:W1_VAR_D1}
\max_{W_1}\,\VAR[X|Y_1,W_1] = \min(\VAR[X|Y_1],\, D_1)
\end{align}
where the first term in (\ref{eq:W1_VAR_D1}) follows from the non-negativity of $I(X;W_1|Y_1)$, and the second one from the distortion constraint on $\hat{X}_1$
\begin{align}
\VAR[X|Y_1,W_1] = \E\bigl[\bigl(X-\hat{X}_1(Y_1,W_1)\bigr)^2\bigr] \leq D_1.
\end{align}

Next, we consider the construction of $W_1,W_2$ that achieves the rate--distortion function, namely, jointly Gaussian random variables with conditional variances that satisfy (\ref{eq:W2_VAR_D2}), (\ref{eq:W1_VAR_D1}).
We construct the optimal distribution $W_1^*,W_2^*$ as follows:
\begin{align}
\label{eq:W1_a1_X_N1}
W_1^* &= \sqrt{a_1} X + N_1\\
\label{eq:W2_a2_X_N2}
W_2^* &= \sqrt{a_2} X + N_2
\end{align}
where $N_i\sim\mbox{ i.i.d. }\mathcal{N}(0,1)$, $i=1,2$, is independent from $X,Y_1,Y_2$, and $a_1,a_2$ are nonnegative scalars whose values are to be specified.
For notational convenience, we define
\begin{align}
\label{eq:V_1_def_W_1_VAR}
V_1 &\triangleq \VAR[X|Y_1,W_1^*]\\
\label{eq:V_1_min_D_1}
&= \min\bigl((\sigma_X^{-2}+s_1)^{-1},\, D_1\bigr)
\end{align}
where (\ref{eq:V_1_min_D_1}) follows from (\ref{eq:W1_VAR_D1}).
Substitute (\ref{eq:W1_a1_X_N1}) in (\ref{eq:V_1_def_W_1_VAR}), and $a_1$ evaluates to
\begin{align}
\label{a1_V_s}
a_1 &= V_1^{-1} -\sigma_X^{-2} - s_1.
\end{align}
Similarly, to identify the optimal $W_2^*$, we define
\begin{align}
\label{eq:V_2_def_W_2_VAR}
V_2 &\triangleq \VAR[X|Y_2,W_1^*,W_2^*]\\
\label{eq:V_2_min_D_2}
&= \min\bigl((V_1^{-1}+s_2-s_1)^{-1},\, D_2\bigr)
\end{align}
which follows from (\ref{eq:W2_VAR_D2}), (\ref{eq:VAR_XY2W1_XY1W1_s}).
Substitute (\ref{eq:W2_a2_X_N2}) in (\ref{eq:V_2_def_W_2_VAR}), and $a_2$ evaluates to
\begin{align}
\label{a2_V_s}
a_2 &= V_2^{-1} -V_1^{-1} -(s_2-s_1).
\end{align}

To provide an interpretation regarding the source encoding rates under different fading states of the side-information channel, we introduce the notations
\begin{align}
R_1 &\triangleq I(X;W_1^*|Y_1)\\
\label{eq:R1_log_V1}
&= \frac{1}{2}\log\frac{(\sigma_X^{-2}+s_1)^{-1}}{V_1}\\
R_2 &\triangleq I(X;W_2^*|Y_2,W_1^*)\\
\label{eq:R2_log_V2}
&= \frac{1}{2}\log\frac{(V_1^{-1}+s_2-s_1)^{-1}}{V_2}
\end{align}
where (\ref{eq:R1_log_V1}), (\ref{eq:R2_log_V2}) follow from expanding the mutual information expressions applying (\ref{a1_V_s}), (\ref{a2_V_s}).
We interpret $R_1$ as the rate of a source coding base layer that describes $X$ when the side-information quality is that of $Y_1$ or better.
On the other hand, $R_2$ is the rate of a top layer that describes $X$ only when the decoder has the better side information $Y_2$.
Finally, we substitute (\ref{eq:R1_log_V1}), (\ref{eq:R2_log_V2}) in (\ref{eq:R_HB_D1D2}) to obtain the two-layer Heegard--Berger rate--distortion function
\begin{align}
R_{\HB}(D_1,D_2) &= R_1 + R_2\\
\label{eq:R_HB_V1_V2}
&= -\frac{1}{2}\log(\sigma_X^{-2}+s_1)-\frac{1}{2}\log V_2 -\frac{1}{2}\log\bigl(1+(s_2-s_1)V_1\bigr)
\end{align}
where $V_1, V_2$ are as defined in (\ref{eq:V_1_def_W_1_VAR}), (\ref{eq:V_2_def_W_2_VAR}) above.
The derivation of (\ref{eq:R_HB_V1_V2}) depends on the side information only through the marginals $p(y_i|x)$'s; therefore, the rate--distortion function applies as well to the stochastically degraded side information $Y'_M,\dotsc,Y'_1$.

\subsection{Optimal Distortion Trade-off and Rate Allocation}
\label{sec:opt_dis_troff_rate_alloc}

\begin{figure}
  \centering
  \includegraphics*[width=8cm]{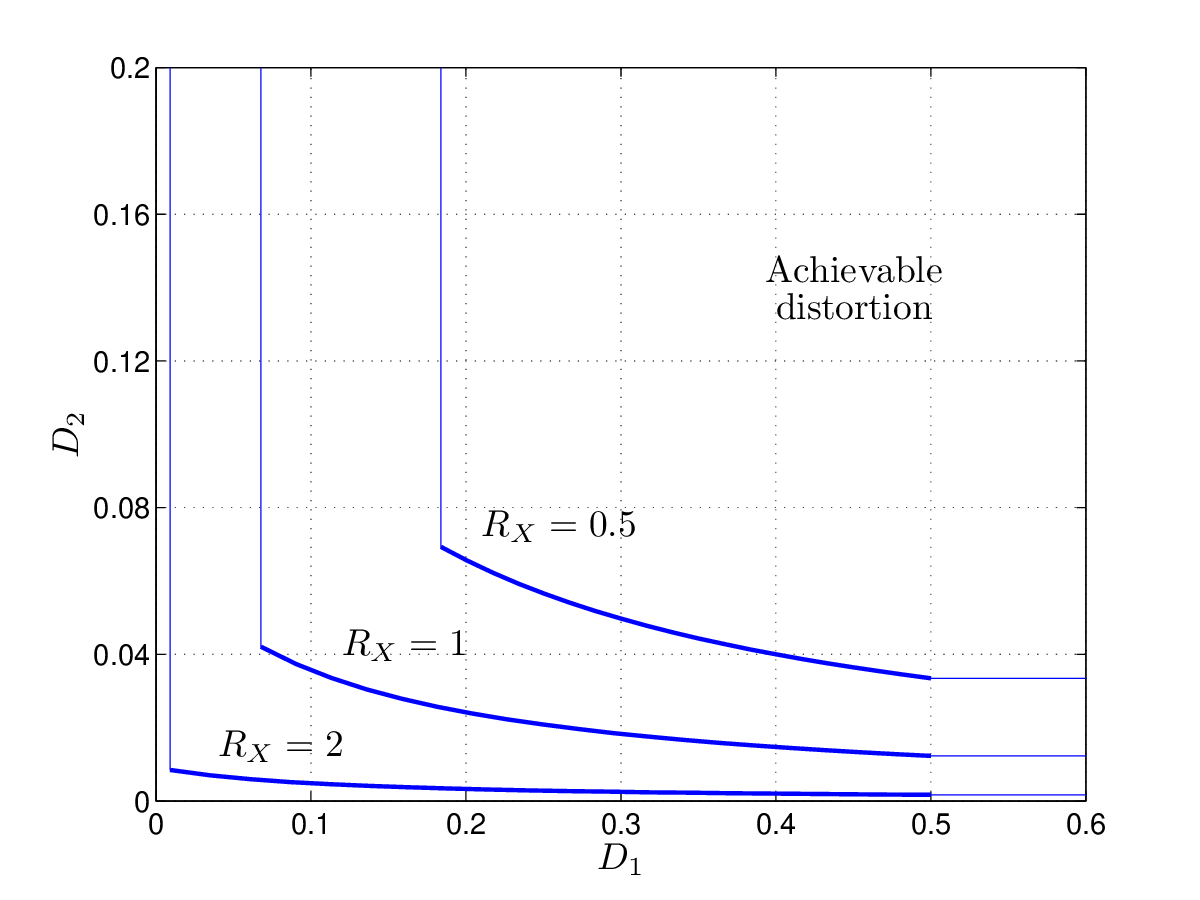}
  \caption{Achievable distortion under different values of the encoding rate constraint $R_X$. For each $R_X$, the Pareto optimal trade-off curve is shown in bold.}
  \label{fig:dist_region}
\end{figure}

Under a source-coding rate constraint of $R_X$, the Heegard--Berger feasible distortion region is described by
\begin{align}
\mathcal{D}(R_X) &\triangleq \{ (D_1,D_2) \st R_{\HB}(D_1,D_2) \leq R_X \}.
\end{align}
The distortion regions under different values of $R_X$ are illustrated in Fig.~\ref{fig:dist_region}.
Setting $R_{\HB}(D_1,D_2)$ $=$ $R_X$, the dominant boundary of $\{(D_1,D_2)\}$ defines the Pareto optimal trade-off curve (shown in bold in Fig.~\ref{fig:dist_region}) between the two distortion constraints on $\hat{X}_1$ and $\hat{X}_2$, which is given by
\begin{align}
\label{eq:D2_D1_Pareto_opt}
D_2 &= \bigl[e^{2R_X}(\sigma_X^{-2}+s_1)\bigl(1+(s_2-s_1)D_1\bigr)\bigr]^{-1}
\end{align}
over the interval
\begin{align}
\bigl(e^{2R_X}(\sigma_X^{-2}+s_1)\bigr)^{-1} \leq D_1 \leq (\sigma_X^{-2}+s_1)^{-1}.
\end{align}
We find the optimal operating point on the Pareto curve to minimize the expected distortion
\begin{align}
\label{eq:ED_min_D1D2}
\E[D]^* = \min_{D_1,D_2\,:\: R_{\HB}(D_1,D_2)\leq R_X} p_1 D_1 + p_2 D_2.
\end{align}
In Section~\ref{sec:mult_disc_fading}, it is shown that the above minimization is a convex optimization problem.
Hence the Karush--Kuhn--Tucker (KKT) conditions are necessary and sufficient for optimality.
Moreover, $\mathcal{D}(R_X)$, being the sublevel set of a convex function, is a convex set.
After substituting (\ref{eq:D2_D1_Pareto_opt}) in (\ref{eq:ED_min_D1D2}), from the KKT optimality conditions, we obtain the optimal base layer distortion
\begin{align}
D_1^* &= \bigl(D_1^*\star)_{[D_1^-,\,D_1^+]}
\end{align}
where $(x)_{[a,\,b]}$ denotes the projection
\begin{align}
(x)_{[a,\,b]} &\triangleq \min\bigl(\max(a,\,x),\,b\bigr)
\end{align}
and the distortion and its boundaries are given by
\begin{align}
D_1^- &\triangleq (e^{2R_X}(\sigma_X^{-2}+s_1)\bigr)^{-1}\\
D_1^\star &\triangleq \frac{1}{s_2-s_1}
\Bigl[\Bigl(e^{2R_X}\frac{\sigma_X^{-2}+s_1}{s_2-s_1}\frac{p_1}{p_2}\Bigr)^{-1/2}-1\Bigr]\\
D_1^+ &\triangleq (\sigma_X^{-2}+s_1)^{-1}.
\end{align}
The optimal top layer distortion $D_2^*$ is given by
\begin{align}
D_2^* &= \bigl(D_2^\star\bigr)_{[D_2^-,\,D_2^+]}
\end{align}
where
\begin{align}
D_2^- &\triangleq (e^{2R_X}(\sigma_X^{-2}+s_2)\bigr)^{-1}\\
D_2^\star &\triangleq \bigl(e^{2R_X}(\sigma_X^{-2}+s_1)(s_2-s_1)p_2/p_1\bigr)^{-1/2}\\
D_2^+ &\triangleq \bigl(e^{2R_X}(\sigma_X^{-2}+s_1)+s_2-s_1\bigr)^{-1}.
\end{align}
The corresponding optimal rate allocation $R_1^*,R_2^*$ can be found as given in (\ref{eq:R1_log_V1}), (\ref{eq:R2_log_V2}).

\begin{figure}
  \centering
  \includegraphics*[width=8cm]{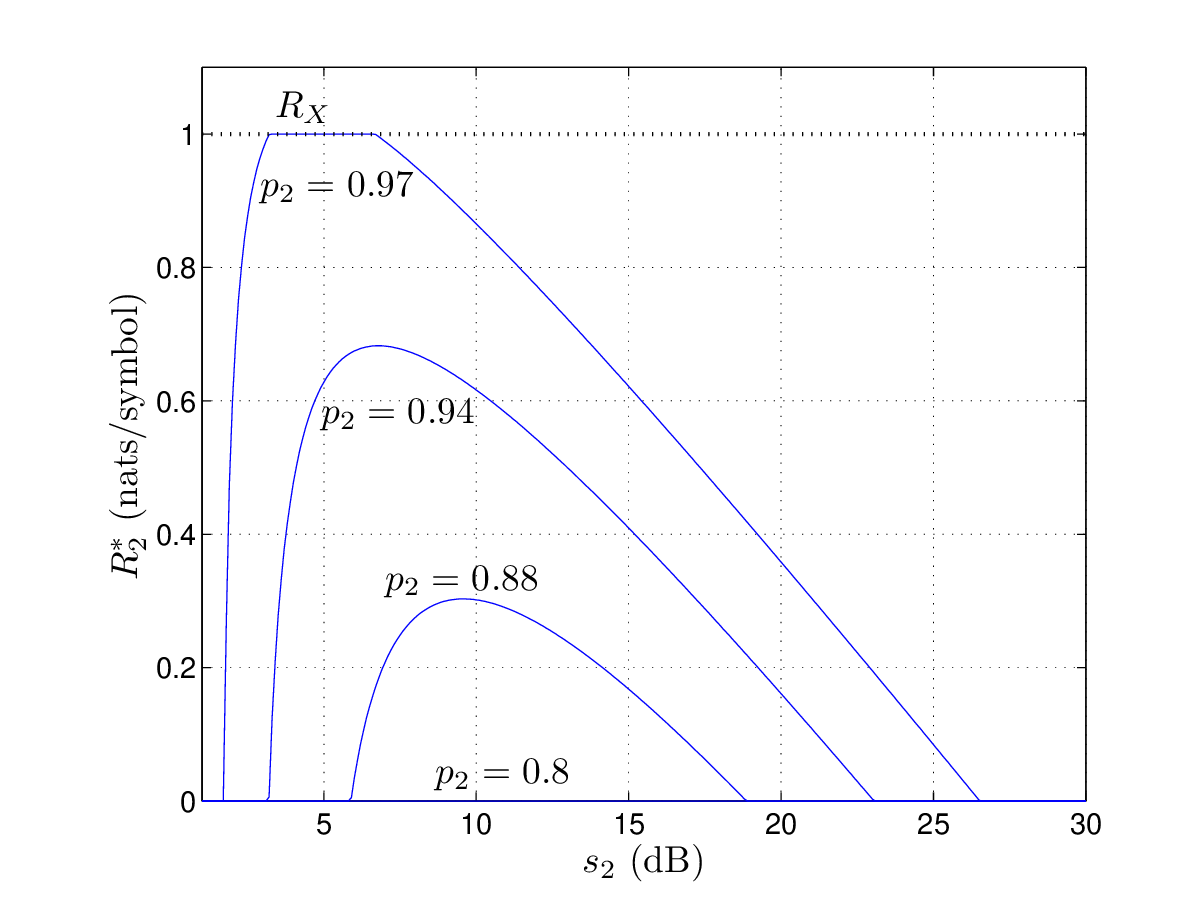}
  \caption{Optimal rate allocation that minimizes expected distortion. The side-information channel has two discrete fading states ($s_1= 0\,\dB$).}
  \label{fig:opt_p_s_R2}
\end{figure}

\begin{figure}
  \centering
  \includegraphics*[width=8cm]{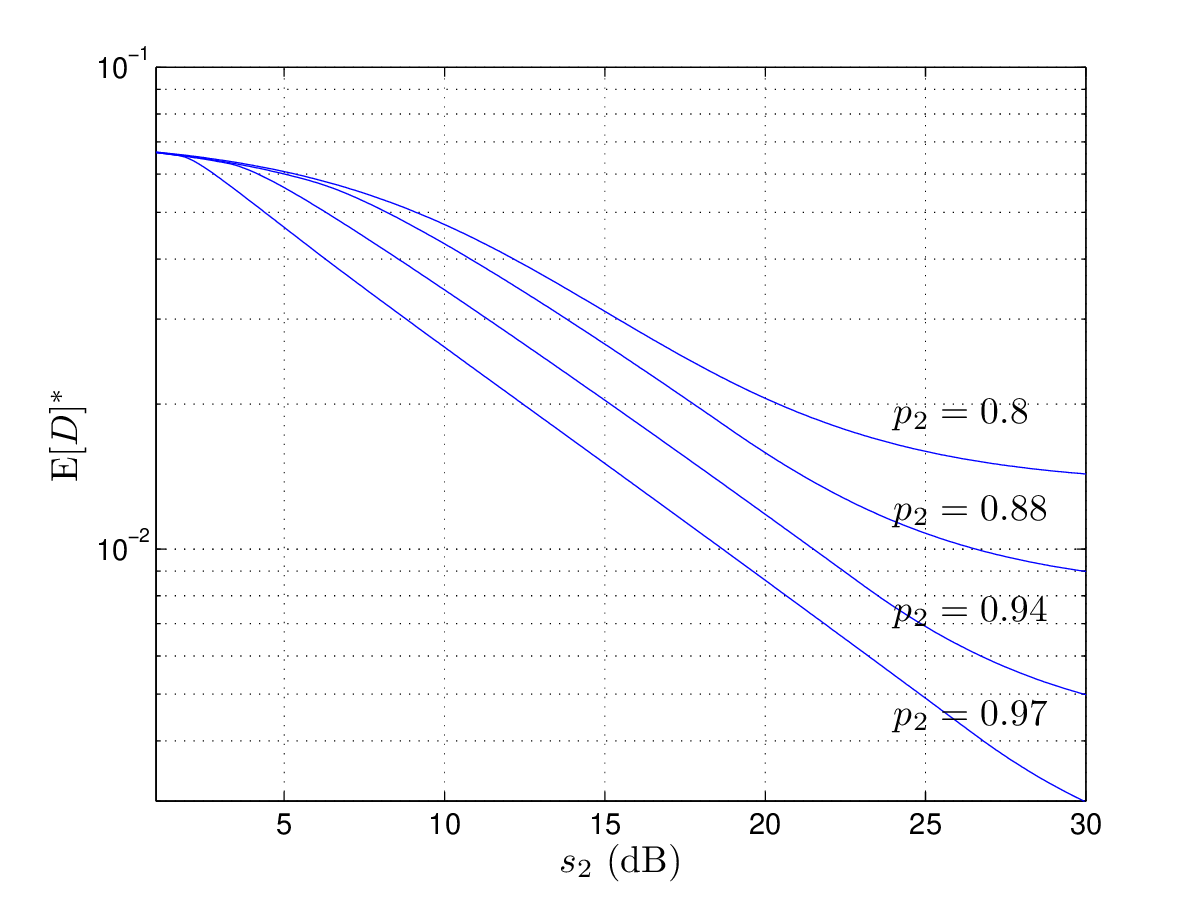}
  \caption{Minimum expected distortion under optimal rate allocation. The side-information channel has two discrete fading states ($s_1= 0\,\dB$).}
  \label{fig:opt_p_s_ED}
\end{figure}

The optimal rate allocation and the corresponding minimum expected distortion are plotted in Fig.~\ref{fig:opt_p_s_R2} and Fig.~\ref{fig:opt_p_s_ED}, respectively, for $R_X=1$, $\sigma_X^2=1$, and $s_1=0\,\dB$.
Note that $R_2^*$, the rate allocated to the top layer, is not monotonic with the side-information channel condition. As fading state $s_2$ improves, $R_2^*$ increases to take advantage of the better side-information quality.
However, when $s_2$ is large, $R_2^*$ begins to decline as the expected distortion is dominated by the worst fading state.
In addition, the optimal rate allocation is heavily skewed towards the lower layer: $R_2^* > 0$ only when $p_2$ is large.

\subsection{Multiple Discrete Fading States}
\label{sec:mult_disc_fading}

The rate--distortion function (\ref{eq:R_HB_V1_V2}) extends directly to the case when the side-information channel has multiple discrete fading states:
$S=s_i$ with probability $p_i$, where $i=1,\dotsc,M$, with $0\leq s_1 < \dotsb < s_M$, and $M>2$.
The Heegard--Berger rate--distortion function for $M>2$ can be characterized by a similar representation as that given in (\ref{eq:R_HB_V1_V2}) for $M=2$.
Specifically, we construct the optimal distribution for the auxiliary random variable $W_i^*$'s to be given by
\begin{align}
W_i^* = \sqrt{a_i} X + N_i,\qquad i=1,\dotsc,M
\end{align}
where $N_i\sim\mbox{ i.i.d. }\mathcal{N}(0,1)$, and $a_i$'s are nonnegative scalars whose values are to be specified.
The rate of the $i\,$th layer is
\begin{align}
R_i &\triangleq I(X;W_i^*|Y_i,W_1^*,\dotsc,W_{i-1}^*)\\
\label{eq:Ri_Vi1_Vi}
&= \frac{1}{2}\log\frac{(V_{i-1}^{-1}+s_i-s_{i-1})^{-1}}{V_i}
\end{align}
where
\begin{align}
\label{eq:Vi_VAR_X_Yi_W1i}
V_i &\triangleq \VAR[X|Y_i,W_1^*,\dotsc,W_i^*]\\
\label{eq:Vi_min_Di_Vi1_s}
&= \min\bigl((V_{i-1}^{-1}+s_i-s_{i-1})^{-1},\, D_i\bigr)
\end{align}
and $s_0 \triangleq 0$, $V_0 \triangleq \sigma_X^2$ for convenience in notations.
In the above, (\ref{eq:Vi_min_Di_Vi1_s}) follows from the non-negativity of $I(X;W_i|Y_i,W_1,\dotsc,W_{i-1})$ and the distortion constraint (\ref{eq:E_d_Xi_Xhi_Di}).
The $a_i$ that achieves (\ref{eq:Vi_VAR_X_Yi_W1i}) is determined from (\ref{eq:Vi_min_Di_Vi1_s}), which evaluates to
\begin{align}
a_i = V_i^{-1} - V_{i-1}^{-1} - (s_i-s_{i-1}).
\end{align}
As $R_{\HB}(\mathbf{D}) = \sum_{i=1}^M R_i$, we substitute (\ref{eq:Ri_Vi1_Vi}) in (\ref{eq:R_HB_D}) to obtain the rate--distortion function
\begin{align}
\label{eq:R_HB_Vi}
R_{\HB}(\mathbf{D}) &= -\frac{1}{2}\log(\sigma_X^{-2}+s_1)-\frac{1}{2}\log V_M-\frac{1}{2}\sum_{i=1}^{M-1}\log\bigl(1+(s_{i+1}-s_i) V_i\bigr)
\end{align}
where the $V_i$'s are as given in (\ref{eq:Vi_min_Di_Vi1_s}).

Under multiple fading states, however, a closed-form expression for the minimum expected distortion $\E[D]^*$ does not appear analytically tractable.
Nevertheless, the expected distortion minimization in (\ref{eq:E_D_opt_RD}) can be formulated as the following convex optimization problem:
\begin{align}
\label{eq:cvx_DV_J}
&\text{minimize}\quad J(D_1,\dotsc,D_M)\\
\label{eq:cvx_over_D1_DM_V1_VM_Rpp}
&\text{over}\quad D_1,\dotsc,D_M,\, V_1,\dotsc,V_M \in \mathds{R}_{++}\\
&\text{subject to}\notag\\
\label{eq:cvx_DV_log_V_Rx}
&\quad -\frac{1}{2}\log(\sigma_X^{-2}+s_1) -\frac{1}{2}\log V_M -\frac{1}{2}\sum_{i=1}^{M-1}\log\bigl(1+(s_{i+1}-s_i) V_i\bigr) \leq R_X\\
\label{eq:cvx_DV_Vi_si}
&\quad V_i \leq (V_{i-1}^{-1}+s_i-s_{i-1})^{-1},\quad i=1,\dotsc,M\\
\label{eq:cvx_DV_Vi_Di}
&\quad V_i \leq D_i,\quad i=1,\dotsc,M
\end{align}
where $\mathds{R}_{++}$ denotes the set of positive real numbers.
In (\ref{eq:cvx_DV_J}) above, the cost function $J(\cdot)$ may be any arbitrary function that is convex in $D_1,\dotsc,D_M$.
The constraint (\ref{eq:cvx_DV_log_V_Rx}) prescribes the feasible Heegard--Berger distortion region under the source-coding rate constraint $R_X$.
The constraints (\ref{eq:cvx_DV_Vi_si}) and (\ref{eq:cvx_DV_Vi_Di}) derive from writing out the two upper bounds for each $V_i$, as described in (\ref{eq:Vi_min_Di_Vi1_s}), as two separate inequality constraints.
The equality in (\ref{eq:Vi_min_Di_Vi1_s}) may be written as inequality constraints since there is an optimal solution where for each~$i$ at least one of (\ref{eq:cvx_DV_Vi_si}) or (\ref{eq:cvx_DV_Vi_Di}) is tight.
Specifically, the left-hand side of the Heegard--Berger constraint in (\ref{eq:cvx_DV_log_V_Rx}) is monotonically decreasing in $V_i$'s.
Hence for a given optimal $\{V_i^*, D_i^*\}$, if neither (\ref{eq:cvx_DV_Vi_si}) nor (\ref{eq:cvx_DV_Vi_Di}) is tight, $V_i^*$ may be increased to strictly enlarge the feasible set of $\{D_1,\dotsc,D_M,\, V_1,\dotsc,V_M\} \backslash \{D_i,V_i\}$.

\begin{prop}
\label{prop:cost_min_convex}
The minimization given in (\ref{eq:cvx_DV_J})--(\ref{eq:cvx_DV_Vi_Di}) is a convex optimization problem.
\end{prop}

See Appendix~\ref{sec:proof_cost_min_convex} for the proof of Proposition~\ref{prop:cost_min_convex}.
Convexity implies that a local optimum is globally optimal,
and its solution can be efficiently computed by standard convex optimization numerical techniques,
for instance, by the interior-point method \cite{renegar01:math_ipm_cvxopt, boyd04:convex_opt}.
Moreover, the optimization problem (\ref{eq:cvx_DV_J})--(\ref{eq:cvx_DV_Vi_Di}) has $2M$ variables and $2M+1$ inequality constraints, which are linear in the number of side-information channel fading states $M$.

In the case where the cost function $J(D_1,\dotsc,D_M)$ is non-decreasing in each component $D_i$, the constraints (\ref{eq:cvx_DV_Vi_Di}) may be taken as tight: if at their optimal values $V_i^* < D_i^*$, then $D_i^*$ may be decreased without violating feasibility nor increasing the cost function.
In particular, in the remainder of the paper, we consider minimizing the expected distortion:
\begin{align}
\label{eq:J_D_expected_distortion}
J(\mathbf{D}) = \E[D] = \sum_{i=1}^M p_i D_i.
\end{align}
In this case, the optimization problem can be specified more compactly as
\begin{align}
\label{eq:cvx_ED_pD}
&\text{minimize}\quad p_1 D_1 + \dotsb + p_M D_M\\
\label{eq:cvx_over_D1_DM}
&\text{over}\quad D_1,\dotsc,D_M\\
&\text{subject to}\notag\\
\label{eq:cvx_ED_D_RX}
&\quad -\frac{1}{2}\log(\sigma_X^{-2}+s_1) -\frac{1}{2}\log D_M -\frac{1}{2}\sum_{i=1}^{M-1}\log\bigl(1+(s_{i+1}-s_i) D_i\bigr) \leq R_X\\
\label{eq:cvx_ED_Di_si}
&\quad D_i \leq (D_{i-1}^{-1}+s_i-s_{i-1})^{-1},\quad i=1,\dotsc,M
\end{align}
where in (\ref{eq:cvx_ED_Di_si}) similarly $D_0 \triangleq \sigma_X^2$.
For convenience in stating the optimization problem, in (\ref{eq:cvx_ED_D_RX}),
and in the remainder of the paper,
$\log$ refers to the extended-value logarithmic function,
where it takes on the value $-\infty$ for non-positive arguments.
Then the feasibility constraints (\ref{eq:cvx_ED_D_RX}), (\ref{eq:cvx_ED_Di_si})
imply $D_i>0$, $i=1,\dotsc,M$,
and the domain qualification $\{D_i\in\mathds{R_{++}}\}$ is thus omitted from (\ref{eq:cvx_over_D1_DM}).
The positivity of $D_i$ can be shown as follows.
Note that (\ref{eq:cvx_ED_D_RX}) implies
\begin{align}
\label{eq:D_M_gt_0}
D_M &> 0\\
\label{eq:s_D_i_gt_0}
1+(s_{i+1}-s_i)D_i &>0, \qquad i=1,\dotsc,M-1.
\end{align}
Consider $i=M-1$.
Suppose $D_{M-1} < 0$, then (\ref{eq:s_D_i_gt_0}) rearranges to
$(D_{M-1}^{-1} + s_M-s_{M-1})^{-1} < 0$,
which contradicts (\ref{eq:cvx_ED_Di_si}), (\ref{eq:D_M_gt_0}):
$(D_{M-1}^{-1} + s_M-s_{M-1})^{-1} \geq D_M > 0$.
Next, suppose $D_{M-1} = 0$.
Applying $\lim_{D_{M-1}\rightarrow0^+}$ on (\ref{eq:cvx_ED_Di_si}),
the inequality becomes: $D_M \leq 0$, which contradicts (\ref{eq:D_M_gt_0}).
Therefore, $D_{M-1}>0$, and similar arguments apply for $D_{M-2},\dotsc,D_1$.

\section{Rate Allocation Under Different Fading Distributions}
\label{sec:rate_alloc_fading}

\begin{figure}
  \centering
  \includegraphics*[width=8cm]{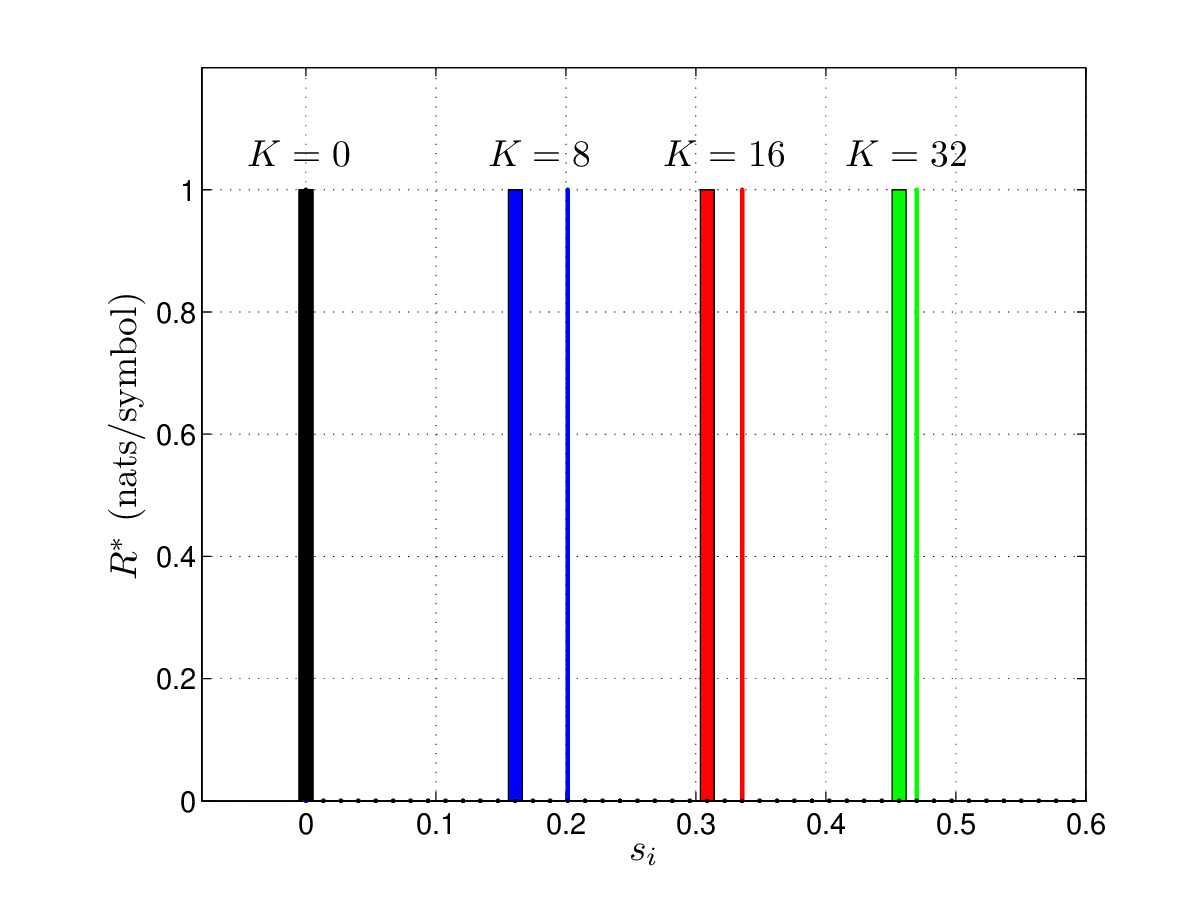}
  \caption{Optimal rate allocation that minimizes the expected distortion $\E[D]$. The rate allocation corresponding to Rician fading is shown in bars, and the one corresponding to Nakagami fading with $m=(K+1)^2/(2K+1)$ is shown in lines. In each case, the optimal rate allocation is concentrated at a single layer. ($R_X=1$, $\bar{S}=1$, $s_M=2\bar{S}$, $R_X=1$, $\sigma_X^2=1$, $M=150$).}
  \label{fig:cmp_fig_ED_Rice_Naka_K}
\end{figure}

\begin{figure}
  \centering
  \includegraphics*[width=8cm]{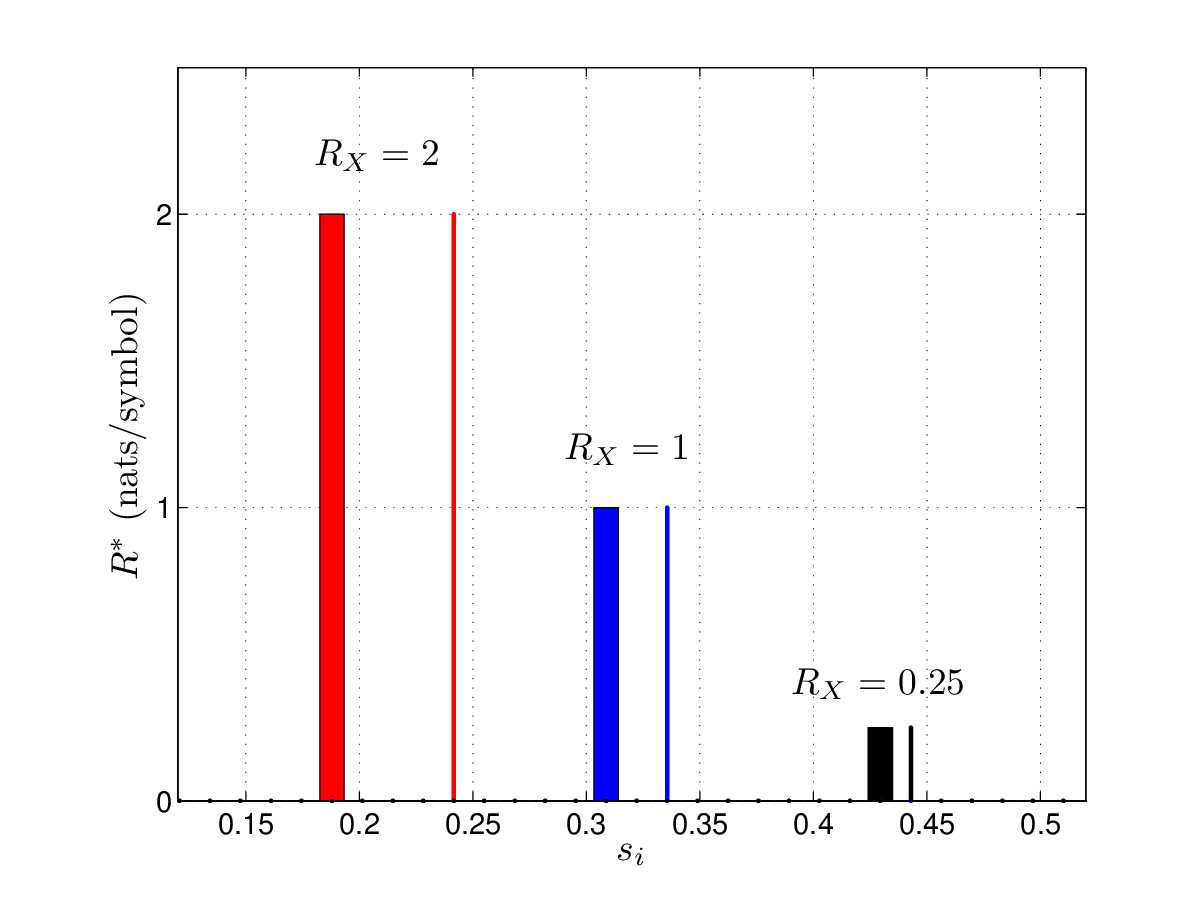}
  \caption{Optimal rate allocation that minimizes the expected distortion $\E[D]$ with $K=16$ under different values of $R_X$ (the other parameters are the same as those in Fig.~\ref{fig:cmp_fig_ED_Rice_Naka_K}). In each case, the optimal rate allocation is concentrated at a single layer.}
  \label{fig:cmp_fig_ED_Rice_Naka_Rx}
\end{figure}

In this section, we apply the optimization framework developed in Section~\ref{sec:mult_disc_fading}, and study the optimal rate allocation when the side-information channel is subject to different fading distributions.
We first consider the scenario when the side-information channel experiences Rician fading, the pdf of which is given by
\begin{align}
\label{eq:fRs_Rician_pdf}
f_{\mathrm{C}}(s) = \frac{(1+K)e^{-K}}{\bar{S}} \exp\Bigl(-\frac{(1+K)s}{\bar{S}}\Bigr) I_0\biggl(2\sqrt{\frac{K(1+K)s}{\bar{S}}}\biggr),
\quad s\geq 0
\end{align}
where $I_0(\cdot)$ is the modified Bessel function of zeroth order, and $\bar{S}$ is the mean channel power gain.
The Rician $K$-factor represents the power ratio of the line-of-sight (LOS) component to the non-LOS components.
Specifically, (\ref{eq:fRs_Rician_pdf}) reduces to Rayleigh fading for $K=0$, and to no fading (i.e., constant channel power gain of $\bar{S}$) for $K=\infty$.
We discretize the channel fading pdf into $M$ states
\begin{align}
p_i &= \Prob\{\text{Side information channel state $s_i$ is realized}\}\\
&= \int_{s_i}^{s_{i+1}} f(s) \,ds ,\quad i=1,\dotsc,M
\end{align}
where we truncate the pdf at $s_M$.
The quantized channel power gains are evenly spaced: $s_i = (i-1)s_M/(M-1)$, $i=1,\dotsc,M$, and $s_{M+1} \triangleq \infty$.
In the numerical experiments, the convex optimization problems are solved using the primal-dual interior-point algorithm described in \cite[Section~11.7]{boyd04:convex_opt}.
The optimal rate allocation that minimizes the expected distortion $\E[D]$ is shown in Fig.~\ref{fig:cmp_fig_ED_Rice_Naka_K} and Fig.~\ref{fig:cmp_fig_ED_Rice_Naka_Rx}, respectively, for different values of $K$ and $R_X$ with $M=150$.
For comparison, we also show in the figures the optimal rate allocation under Nakagami fading with the pdf
\begin{align}
\label{eq:fNs_Nakagami_pdf}
f_{\mathrm{N}}(s) = \frac{(m/\bar{S})^m s^{m-1} e^{-ms/\bar{S}}}{\Gamma(m)}, \quad s\geq 0
\end{align}
where $\Gamma(\cdot)$ is the gamma function.
In Fig.~\ref{fig:cmp_fig_ED_Rice_Naka_K} and Fig.~\ref{fig:cmp_fig_ED_Rice_Naka_Rx}, the Nakagami parameter $m$ is set to be: $m=(K+1)^2/(2K+1)$, under which the Nakagami distribution (\ref{eq:fNs_Nakagami_pdf}) is commonly used to approximate the Rician distribution in (\ref{eq:fRs_Rician_pdf}) \cite{stuber00:prin_mob_comm}.

\begin{figure}
  \centering
  \includegraphics*[width=8cm]{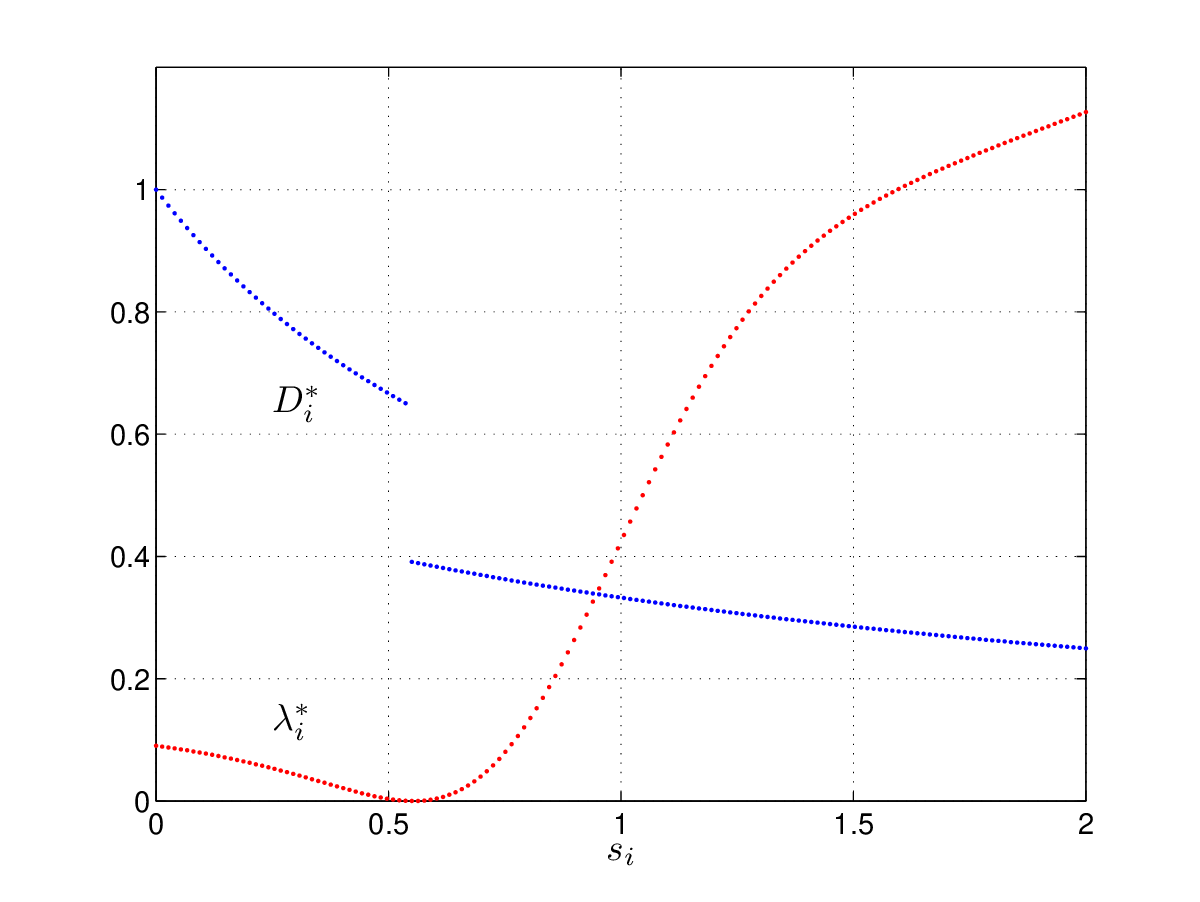}
  \caption{Optimal primal and dual variables in the expected distortion minimization under Rician fading with $K=32$, $\bar{S}=1$, $R_X=0.25$, $\sigma_X^2=1$.
  The rate constraint (\ref{eq:cvx_ED_D_RX}) is tight, and $\mu^* \approx 0.56$.}
  \label{fig:D_lamb_Rician_co}
\end{figure}

In each case of the numerical results, it is observed that the optimal rate allocation is concentrated at a \emph{single} layer, i.e., $R_i^* = R_X$ for some $i=i^*$ at $s_{i^*}$, while $R_i^* = 0$ for all other $i\neq i^*$.
The optimal primal and dual variables $D_i^*, \lambda_i^*$ are plotted in Fig.~\ref{fig:D_lamb_Rician_co} for the case of Rician fading with $K=32$, $\bar{S}=1$, $R_X=0.25$, $\sigma_X^2=1$.
In this case, the rate allocation concentrates at $s_{i^*}\approx0.55$, and the complementary slackness condition (\ref{eq:KKT_ED_lamb}) stipulates that the corresponding dual variable be zero: $\lambda_{i^*}=0$.
In Fig.~\ref{fig:cmp_fig_ED_Rice_Naka_K},
under Rayleigh fading ($K=0$), the optimal rate allocation concentrates at the base layer (i.e., $s_{i^*}=0$) of the source code.
In the case where the side-information channel has a prominent LOS component, i.e., when $K$ is large, $s_{i^*}$ increases accordingly as the channel distribution is more concentrated around $\bar{S}$.
On the other hand, a large source-coding rate $R_X$ decreases $s_{i^*}$, which implies that it is less beneficial to be opportunistic to target possible good channel conditions when $R_X$ is large.
Moreover, for each $\bar{S}$, Nakagami fading results in a higher $s_{i^*}$ than its corresponding Rician fading distribution.

\begin{figure}
  \centering
  \includegraphics*[width=8cm]{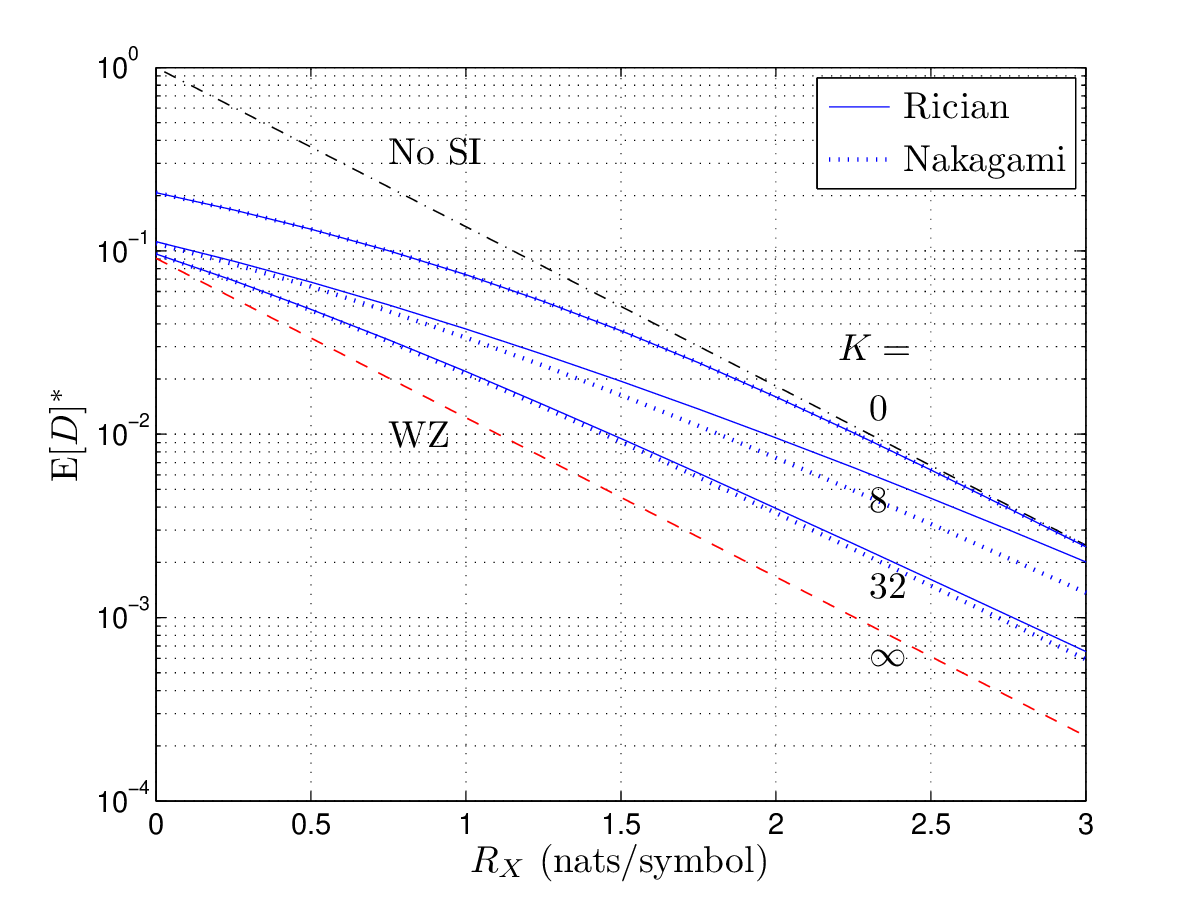}
  \caption{Minimum expected distortion. The dash-dot line corresponds to the rate--distortion function with no side information (No~SI). The dashed line ($K=\infty$, i.e., the side-information channel has no fading) corresponds to the Wyner--Ziv (WZ) rate--distortion function. ($\bar{S}=10$, $s_M=2\bar{S}$, $\sigma_X^2=1$, $M=150$.)}
  \label{fig:cmp_fig_ED_Rician_Rx_K}
\end{figure}

The minimum expectation $\E[D]^*$ that corresponds to the optimal rate allocation is shown in Fig.~\ref{fig:cmp_fig_ED_Rician_Rx_K}.
For comparison, along with $\E[D]^*$, in Fig.~\ref{fig:cmp_fig_ED_Rician_Rx_K} we also show the distortion under different assumptions on the side information.
When no side information is available, the distortion is given by the rate--distortion function for a Gaussian source \cite{cover91:eoit}
\begin{align}
\label{eq:D_NoSI_Rx}
D_{\NoSI}(R_X) &= \sigma_X^2 e^{-2R_X}.
\end{align}
In the absence of side information, $D_{\NoSI}$ is an upper bound to $\E[D]^*$.
On the other hand, when $K=\infty$, there is no uncertainty in the side-information channel condition with $S=\bar{S}$, and the distortion is given by the Wyner--Ziv \cite{wyner76:rate_dist_side_info} rate--distortion function
\begin{align}
D_{\WZ}(R_X) &= (\sigma_X^{-2}+\bar{S})^{-1} e^{-2R_X}.
\end{align}
In Fig.~\ref{fig:cmp_fig_ED_Rician_Rx_K}, a larger $K$ decreases the expected distortion $\E[D]^*$, and Nakagami fading has a lower $\E[D]^*$ than the corresponding Rician fading distribution.
In addition, when $R_X$ is small, $\E[D]^*$ considerably outperforms $D_{\NoSI}$ where no side information is available, as the reduction in $\VAR[X]$ from the side information at the decoder is significant.
However, when $R_X$ is large, the improvement of $\E[D]^*$ over $D_{\NoSI}$ diminishes, as most of the reduction in $\VAR[X]$ is due to the source-coding rate of $R_X$.

In the following, we make a remark on the \emph{distortion exponent} $\Delta$, defined similarly as given in \cite{laneman05:src_ch_parl_ch}, which characterizes the rate of exponential decay in distortion at asymptotically large encoding rates:
\begin{align}
\label{eq:Delta_def_EDR}
\Delta \triangleq \lim_{R_X\rightarrow\infty} -\frac{\log \E[D(R_X)]^*}{2R_X}
\end{align}
where $R_X$ is the source-coding rate, and $\E[D(R_X)]^*$ is the corresponding minimum expected distortion under $R_X$.
We note that the distortion exponent $\Delta$ does not depend on the fading distribution $f(s)$, since
\begin{align}
D_{\WZ}(R_X) &\leq \E[D(R_X)]^* \leq D_{\NoSI}(R_X)\\
\lim_{R_X\rightarrow\infty} -\frac{\log D_{\WZ}(R_X)}{2R_X} &= \lim_{R_X\rightarrow\infty} -\frac{\log D_{\NoSI}(R_X)}{2R_X} = 1.
\end{align}
Therefore, reducing the side-information channel uncertainty (e.g., via deploying multiple antennas or through channel state information feedback) may reduce the expected distortion $\E[D(R_X)]^*$ at finite $R_X$, but it does not improve performance in the asymptotic regime in terms of the rate of exponential decay in the distortion as a function of the encoding rate $R_X$.

\section{Single-Layer Rate Allocation}
\label{sec:single_layer_rate_alloc}

In the numerical experiments presented in Section~\ref{sec:rate_alloc_fading},
the optimal rate allocation concentrates at a single codeword layer under a wide class of fading distributions in the side-information channel.
Motivated by such observations,
in the following, we consider the single-layer rate allocation:
\begin{align}
\label{eq:single_layer_rate_alloc}
\mathcal{R}_{\bar{s}} &= \{\text{encoding rate $R_X$ at fading state $\bar{s}$}\}
\end{align}
and investigate the conditions under which $\mathcal{R}_{\bar{s}}$ is indeed expected-distortion--minimizing.
First, we consider the case where the side-information channel has infinitely many discrete fading states; based on the discrete fading results, we then consider continuous fading distributions.

\subsection{Infinitely Many Discrete Fading States}
\label{sec:inf_many_disc_fading_states}

Suppose the side-information channel has infinitely many discrete fading states
$\{s_1, s_2, \dotsc\}$
with $0\leq s_1 < s_2 < \cdots$.
Let $p_i$ denote the probability the side-information channel takes on state $s_i$, with $\sum_{i=1}^\infty p_i = 1$.
Let $\mathbf{D} \triangleq [D_1\; D_2\; \dots]^T$ denote the set of distortion variables induced by the rate allocation scheme.
Note that the Heegard--Berger rate--distortion function extends to the set of infinitely many degraded fading states.
We assume the total encoding rate $R_X$ is finite, and hence so is the rate $R_i$ for each fading state.
For admissibility, the random coding argument in \cite[Section~III]{heegard85:rate_dist_si} applies for each of the fading states,
and the converse in \cite[Section~VII]{heegard85:rate_dist_si} depends only on the total encoding rate but not the number of fading states.
Furthermore, it can be shown using the same approach in \cite{tian08:si_scal_src_cod} that the optimality of Gaussian solution for the Gaussian side information Heegard-Berger problem still holds for this scenario.
The expected distortion minimization problem is given by
\begin{align}
\label{eq:inf_disc_ED_pD}
&\text{minimize}\quad \sum_{i=1}^\infty p_i D_i\\
\label{eq:inf_disc_over_D1_DM}
&\text{over}\quad D_1,D_2,\dotsc\\
&\text{subject to}\notag\\
\label{eq:inf_disc_ED_D_RX}
&\quad -\frac{1}{2} \sum_{i=1}^\infty
\Bigl(
\log(D_{i-1}^{-1}+s_i-s_{i-1}) + \log D_i
\Bigr)
\leq R_X\\
\label{eq:inf_disc_ED_Di_si}
&\quad D_i \leq (D_{i-1}^{-1}+s_i-s_{i-1})^{-1},\quad i=1,2,\dotsc.
\end{align}
The rate constraint (\ref{eq:inf_disc_ED_D_RX}) follows from (\ref{eq:Ri_Vi1_Vi}).
Recall that the extended-value logarithm takes on $-\infty$ for non-positive arguments.
The constraint (\ref{eq:inf_disc_ED_D_RX}) then implies $D_i > 0$;
hence the domain $D_i>0$ is not explicitly stated in the optimization problem.
In Appendix~\ref{sec:inf_disc_exp_dist_min}, it is shown that (\ref{eq:inf_disc_ED_pD})--(\ref{eq:inf_disc_ED_Di_si}) is a convex problem,
and its KKT optimality conditions and dual function are given in Appendix~\ref{sec:inf_disc_KKT_opt_dual_fn}.
Next, we show that the single-layer rate allocation (\ref{eq:single_layer_rate_alloc})
is expected-distortion--minimizing in (\ref{eq:inf_disc_ED_pD})--(\ref{eq:inf_disc_ED_Di_si}) under some conditions.

Consider the single-layer rate allocation (\ref{eq:single_layer_rate_alloc})
with $\bar{s}=s_i$ for some $i$.
Relating $\mathcal{R}_{\bar{s}}$ to the induced distortion set (\ref{eq:Ri_Vi1_Vi}), we have
\begin{align}
\label{eq:D_bar_i}
\bar{D}_i &= \begin{cases}
(\sigma_X^{-2} + s_i)^{-1}, & i:\; s_i < \bar{s}\\
\big((\sigma_X^{-2}+\bar{s})e^{2R_X} + s_i - \bar{s}\bigr)^{-1},
& i:\; s_i \geq \bar{s}
\end{cases}
\end{align}
with the single-layer expected distortion given by
\begin{align}
\label{eq:E_D_bar}
\E[\bar{D}] = \sum_{i=1}^\infty p_i \bar{D}_i.
\end{align}
In the notation, the overbar accent is used to represent single-layer rate allocation.
To determine optimality,
we apply the KKT complementary slackness condition (\ref{eq:KKT_ED_lamb}) on (\ref{eq:D_bar_i}) above,
and it stipulates that
\begin{align}
\label{eq:lambda_i_0}
\bar\lambda_i &= 0, \qquad i:\; s_i = \bar{s}.
\end{align}
Next, we apply the KKT gradient conditions (\ref{eq:KKT_ED_dDi}) on (\ref{eq:D_bar_i}), (\ref{eq:lambda_i_0}),
and the dual variables $\bar\lambda_i$ are given as follows:
\begin{align}
\label{eq:lambda_i_1}
\bar\lambda_i &= (\bar{s}-s_i)\frac{\bar\mu}{2}
\frac{\sigma_X^{-2}+s_i}{\sigma_X^{-2}+\bar{s}}
-\sum_{j:\; s_i\leq s_j<\bar{s}}
p_j\biggl(\frac{\sigma_X^{-2}+s_i}{\sigma_X^{-2}+s_j}\biggr)^2,
\qquad i:\; s_i < \bar{s}\\
\label{eq:lambda_i_2}
\begin{split}
\bar\lambda_i &= -(s_i-\bar{s})\frac{\bar\mu}{2}
\biggl(1+\frac{s_i-\bar{s}}{(\sigma_X^{-2}+\bar{s})e^{2R_X}}\biggr)\\
&\qquad+\sum_{j:\;\bar{s}\leq s_j <s_i} p_j\prod_{k:\;s_j\leq s_k <s_i}
\biggl(1+\frac{s_{k+1}-s_k}{(\sigma_X^{-2}+\bar{s})e^{2R_X}+s_k-\bar{s}}\biggr)^2,
\qquad i:\; s_i > \bar{s}.
\end{split}
\end{align}
The dual variable $\bar\mu$ is set such that the rate constraint is tight in (\ref{eq:KKT_ED_mu}):
\begin{align}
\label{eq:mu_bar}
\frac{\bar\mu}{2} &= \sum_{j:\; s_j\geq\bar{s}} p_j
\frac{(\sigma_X^{-2}+\bar{s})e^{2R_X}}
{\bigl((\sigma_X^{-2}+\bar{s})e^{2R_X}+s_j-\bar{s}\bigr)^2}.
\end{align}
The single-layer rate allocation optimality conditions are stated in the following proposition,
and its proof is given in Appendix~\ref{sec:proof_single_layer_lambda}.

\begin{prop}
\label{prop:single_layer_lambda}
In the expected distortion minimization problem where the side-information channel has discrete fading states
$\{s_1, s_2, \dotsc\}$,
the single-layer rate allocation (\ref{eq:single_layer_rate_alloc}) at $\bar{s}=s_i$ for some $i$ is optimal if and only if
\begin{align}
\label{eq:prop_lambda_i_geq_0}
\bar\lambda_i \geq 0, \qquad \forall\; i
\end{align}
where the variables $\bar\lambda_i$ are as given in (\ref{eq:lambda_i_1}), (\ref{eq:lambda_i_2}).
\end{prop}

Approximately, the optimality conditions in (\ref{eq:prop_lambda_i_geq_0}) correspond to
the tail summation in (\ref{eq:mu_bar}) being greater than the partial sum in (\ref{eq:lambda_i_1}) to the left of $\bar{s}$,
but smaller than that in (\ref{eq:lambda_i_2}) to the right of $\bar{s}$, after being weighted by their respective factors.
Next, we consider the case when the side-information channel is described by a continuous fading distribution,
and we show that it admits a remarkably simpler set of sufficient conditions for single-layer rate allocation optimality.

\subsection{Continuous Fading Distributions}
\label{sec:sl_cont_fading_dist}

When the side-information channel has a continuous fading distribution, as defined in Proposition~\ref{prop:ct_single_layer_E_D} below,
the corresponding expected distortion can be derived through a discretization and limiting process.
The expected distortion under continuous fading is stated in the proposition,
and the details of the limiting process can be found in its proof in Appendix~\ref{sec:proof_ct_single_layer_E_D}.

\begin{prop}
\label{prop:ct_single_layer_E_D}
Suppose the side-information channel has a continuous fading distribution, i.e., the cdf $F(s)$ is absolutely continuous with pdf $f(s)=F'(s)$.
We further assume that the pdf $f(s)$ is continuous.
Then, under the single-layer rate allocation $\mathcal{R}_{\bar{s}}$ in (\ref{eq:single_layer_rate_alloc}), the expected distortion is
\begin{align}
\label{eq:E_D_bar_f_s}
\E[\bar{D}] = \int_0^{\bar{s}} \frac{f(s)}{\sigma_X^{-2}+s}\;ds + \int_{\bar{s}}^\infty \frac{f(s)}{(\bar{s}+\sigma_X^{-2})e^{2R_X} + s-\bar{s}} \;ds.
\end{align}
\end{prop}

Next, Proposition~\ref{prop:ct_qconcave_single_opt} describes a sufficient condition for single-layer rate allocation optimality under continuous fading,
and its proof is given in Appendix~\ref{sec:proof_ct_qconcave_single_opt}.
In the proposition,
note that the optimality condition (\ref{eq:lambda_bar_s_geq_0})
is motivated by considering the limiting case of the optimality conditions (\ref{eq:prop_lambda_i_geq_0}) in Proposition~\ref{prop:single_layer_lambda}.

\begin{prop}
\label{prop:ct_qconcave_single_opt}
In Proposition~\ref{prop:ct_single_layer_E_D}, $\mathcal{R}_{\bar{s}}$ is expected-distortion--minimizing if
\begin{align}
\label{eq:lambda_bar_s_geq_0}
\bar\lambda(s) \geq 0, \qquad s\geq0
\end{align}
where
\begin{align}
\label{eq:lambda_bar_s_1}
\bar\lambda(s) &= (\bar{s}-s)\frac{\bar\mu}{2}
\frac{\sigma_X^{-2}+s}{\sigma_X^{-2}+\bar{s}}
- \int_{s}^{\bar{s}}
f(t)\biggl(\frac{\sigma_X^{-2}+s}{\sigma_X^{-2}+t}\biggr)^2 \;dt,
\qquad 0\leq s < \bar{s}\\
\label{eq:lambda_bar_s_2}
\begin{split}
\bar\lambda(s) &= -(s-\bar{s})\frac{\bar\mu}{2}
\biggl(1+\frac{s-\bar{s}}{(\sigma_X^{-2}+\bar{s})e^{2R_X}}\biggr)\\
&\qquad + \int_{\bar{s}}^{s} f(t) \biggl(1+\frac{s-t}{(\sigma_X^{-2}+\bar{s})e^{2R_X}+t-\bar{s}}\biggr)^2
\;dt,
\qquad s \geq \bar{s}
\end{split}
\end{align}
and
\begin{align}
\label{eq:mu_bar_f_s}
\frac{\bar\mu}{2} &= \int_{\bar{s}}^\infty f(s)
\frac{(\sigma_X^{-2}+\bar{s})e^{2R_X}}
{\bigl((\sigma_X^{-2}+\bar{s})e^{2R_X}+s-\bar{s}\bigr)^2}\;ds.
\end{align}
\end{prop}

\subsection{Quasiconcave Probability Density Functions}
\label{sec:sl_qc_pdf}

In the following, we show that fading distributions with continuous, quasiconcave pdfs satisfy the single-layer rate allocation optimality conditions (\ref{eq:lambda_bar_s_geq_0})
in Proposition~\ref{prop:ct_qconcave_single_opt}.
A function $g(x)$ is quasiconcave, or referred to as unimodal, if its superlevel sets $\{x \st g(x)\geq\alpha\}$, for all $\alpha$, are convex.
Notably, most common wireless channel fading distributions have pdfs that are continuous and quasiconcave: e.g., Rayleigh, Rician, Nakagami, and log-normal.

First, we set out the procedure to identify $\bar{s}^*$,
the state at which the encoding rate $R_X$ is to be concentrated.
Let $f(s)$ denote the pdf of the fading distribution of the side-information channel,
and we assume that $f(s)$ is continuous and quasiconcave.
We denote the superlevel set of $f(s)$ by the interval
$[s_\mathrm{a}, s_\mathrm{b}] \triangleq \{s \st f(s)\geq\alpha\}$,
for some nonnegative scalar $\alpha\geq0$.
Specifically, we choose an $\alpha^*$ such that the following relationship holds:
\begin{align}
\label{eq:alpha_f_s_a}
\int_{s_\mathrm{a}(\alpha^*)}^\infty
\frac{f(s)-\alpha^*}
{\bigl((\sigma_X^{-2}+s_\mathrm{a}(\alpha^*))e^{2R_X}+s-s_\mathrm{a}(\alpha^*)\bigr)^2}\;ds = 0.
\end{align}
Note that the left-hand side of (\ref{eq:alpha_f_s_a}) varies continuously from positive to negative as a candidate $\alpha$ ranges from $0$ to $\max f(s)$;
therefore, there exists an $\alpha^*$ that equates the expression to zero.
We set $\bar{s}^*$ to be the left endpoint of the superlevel set induced by $\alpha^*$:
\begin{align}
\label{eq:s_bar_s_a}
\bar{s}^*=s_\mathrm{a}(\alpha^*).
\end{align}
With the single-layer rate allocation target $\bar{s}^*$ properly defined,
the following proposition identifies the class of fading distributions under which this rate allocation is optimal;
its proof is presented in Appendix~\ref{sec:proof_f_s_qc_single_layer}.

\begin{prop}
\label{prop:f_s_qc_single_layer}
Suppose the side-information channel has a continuous fading distribution with pdf $f(s)$.
If $f(s)$ is continuous and quasiconcave,
then the single-layer rate allocation $\mathcal{R}_{\bar{s}^*}$, with $\bar{s}^*$ as given in (\ref{eq:s_bar_s_a}), is expected-distortion--minimizing.
\end{prop}

\begin{figure}
  \centering
  \includegraphics*[width=8cm]{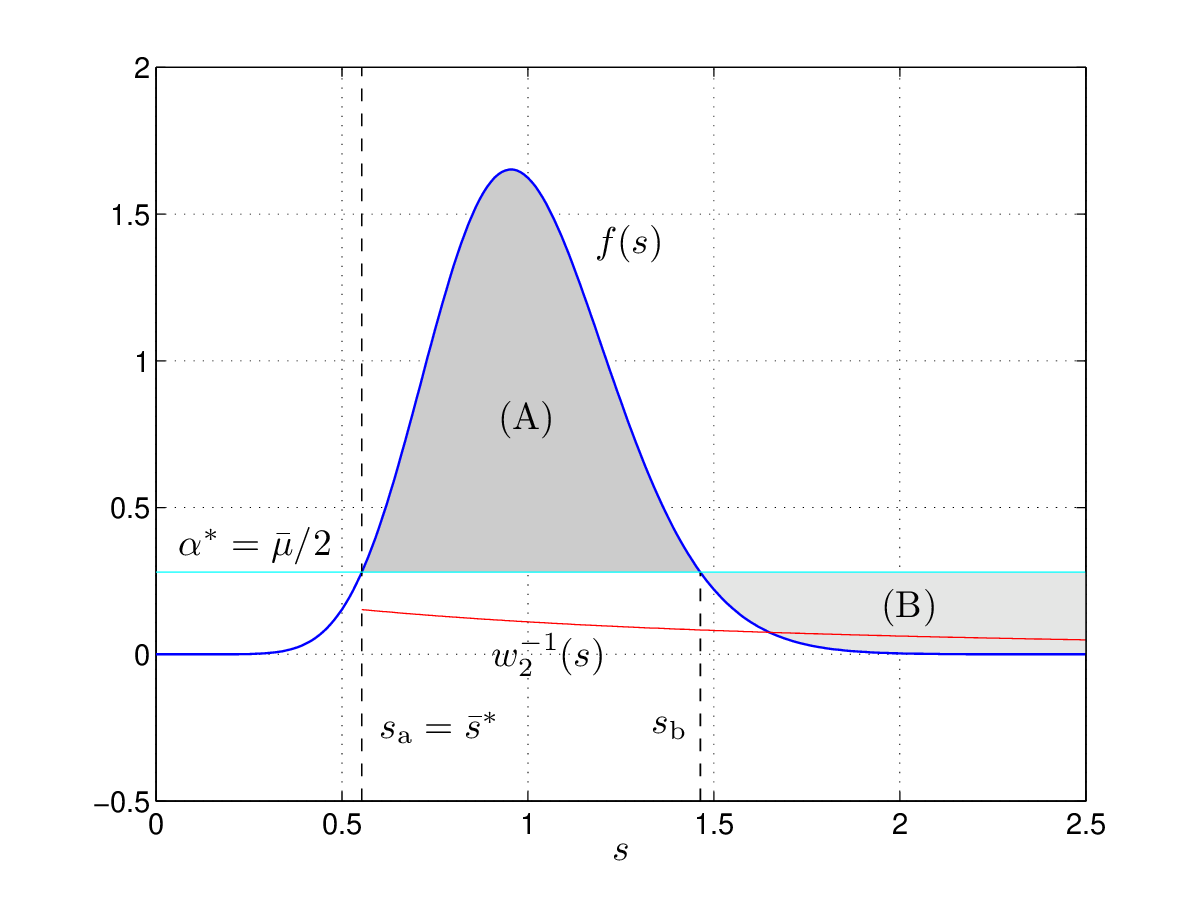}
  \caption{KKT optimality conditions under Rician fading ($K=32$, $R_X=0.25$, $\bar{S}=1$, $\sigma_X^2$=1).
  Note that the area of $(\mathrm{A})$, weighted by $w_2^{-1}(s)$, is equal to the area of $(\mathrm{B})$, weighted by $w_2^{-1}(s)$, where $w_2(s)$ is as defined in (\ref{eq_w_2_s}) in Appendix~\ref{sec:proof_f_s_qc_single_layer}.}
  \label{fig:f_pdf_mu_Rician_co}
\end{figure}

\begin{figure}
  \centering
  \includegraphics*[width=8cm]{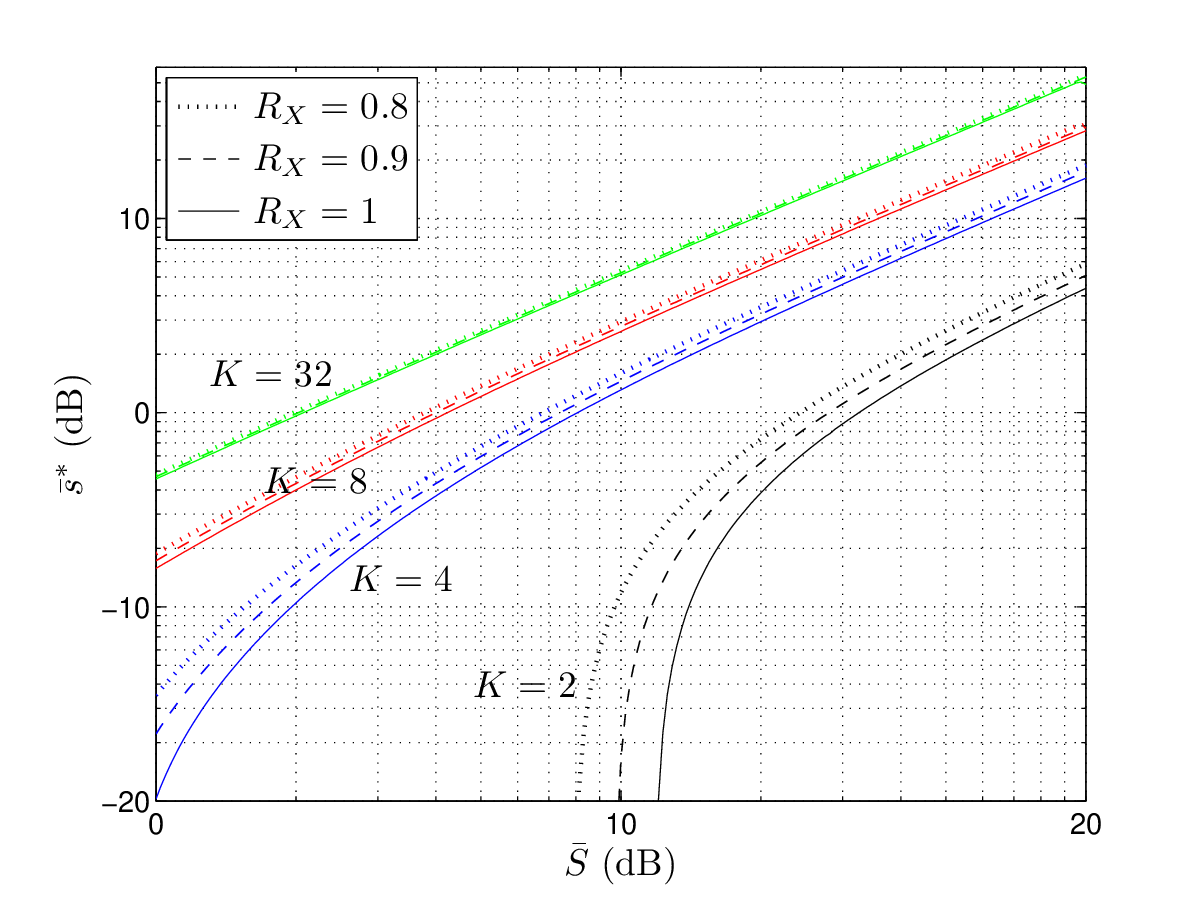}
  \caption{Optimal single-layer rate allocation target $\bar{s}^*$ under Rician fading ($\bar{S}=1$, $\sigma_X^2$=1).}
  \label{fig:ED1_sth_opt_Rician_Rx_K}
\end{figure}

As an example, consider minimizing the expected distortion
under Rician fading ($K=32$, $R_X=0.25$, $\bar{S}=1$, $\sigma_X^2$=1),
for which the KKT optimality conditions are illustrated in Fig.~\ref{fig:f_pdf_mu_Rician_co}.
At optimality, $\alpha^* = \bar\mu/2$.
In the figure, $[s_\mathrm{a},s_\mathrm{b}]$ is the $\bar\mu/2$-superlevel set of $f(s)$,
and the regions between $f(s)$ and $\bar\mu/2$ are shaded and labeled $(\mathrm{A})$ and $(\mathrm{B})$, respectively, for $s_\mathrm{a}\leq s\leq s_\mathrm{b}$ and $s>s_\mathrm{b}$.
The choice of $\alpha^*$ in (\ref{eq:alpha_f_s_a}),
which leads to $\bar{s}^*$ in (\ref{eq:s_bar_s_a}),
corresponds to the area of $(\mathrm{A})$, weighted by $w_2^{-1}(s)$, being equal to the area of $(\mathrm{B})$, weighted by $w_2^{-1}(s)$,
where $w_2(s)$ is as defined in (\ref{eq_w_2_s}) in Appendix~\ref{sec:proof_f_s_qc_single_layer}.
Solving (\ref{eq:alpha_f_s_a}) numerically, the resulting single-layer rate allocation target $\bar{s}^*$
is plotted in Fig.~\ref{fig:ED1_sth_opt_Rician_Rx_K} for different values of $K$ and $R_X$.

As a special case of fading distributions with continuous and quasiconcave pdfs, let us consider Rayleigh fading.
Its pdf is given by
\begin{align}
\label{eq:Ray_pdf}
f_\mathrm{R}(s) &= (1/\bar{S}) e^{s/\bar{S}}, \quad s\geq0
\end{align}
where $\bar{S}$ is the average channel power gain.
Recognizing that any nonempty superlevel set of $f_\mathrm{R}(s)$ begins at $s_\mathrm{a}=0$,
we have the following corollary:
\begin{corr}
\label{corr:Ray_sa_0}
When the side-information channel is under Rayleigh fading (\ref{eq:Ray_pdf}),
the single-layer rate allocation target $\bar{s}^*=0$ is optimal, and the corresponding minimum expected distortion is
\begin{align}
\E[\bar{D}_\mathrm{R}]^* &= \int_0^\infty \frac{(1/\bar{S}) e^{s/\bar{S}}}{\sigma_X^{-2}e^{2R_X}+s}\;ds
= (1/\bar{S})e^{C/\bar{S}}\E_1(C/\bar{S})
\end{align}
where $C\triangleq \sigma_X^{-2}e^{2R_X}$, and $\E_1(\cdot)$ is the exponential integral
$\E_1(x) \triangleq \int_x^\infty \frac{e^{-t}}{t}\;dt$.
\end{corr}
Therefore, under Rayleigh fading, the source-coding scheme does not depend on $\bar{S}$, $R_X$, and $\sigma_X^2$.
It is optimal to concentrate the entire encoding rate $R_X$ at the base layer $\bar{s}^*=0$,
i.e., the source is encoded as if the side information was absent.

\section{Conclusions}
\label{sec:conclu}

We studied the problem of optimal rate allocation and distortion minimization for Gaussian source coding under squared error distortion,
when the uncompressed source symbol is also conveyed over a fading side-information channel.
The encoder knows the fading channel distribution but not its realization.
A layered encoding strategy is used, with each codeword layer targeting the realization of a given fading state.
When the side-information channel has two discrete fading states, we derived closed-form expressions for the optimal rate allocation among the fading states and the corresponding minimum expected distortion.
The optimal rate allocation is conservative: rate is allocated to the higher layer only if the better fading state is highly probable.
For the case of multiple discrete fading states, the minimum expected distortion was shown to be the solution of a convex optimization problem.
We derived an efficient representation for the Heegard--Berger rate--distortion function, under which the number of variables and constraints in the optimization problem is linear in the number of fading states.

Next, we considered single-layer rate allocation, and identified the conditions under which such allocation is expected-distortion--minimizing,
for the respective cases of discrete as well as continuous fading in the side-information channel.
Under continuous fading, we showed that these optimality conditions are satisfied by distributions with continuous and quasiconcave pdfs, e.g., Rayleigh, Rician, Nakagami, and log-normal.
Moreover, for Rayleigh fading, the optimal rate allocation concentrates at the base layer: i.e.,
the source is encoded as if the side information was absent.

In this paper, we focused on fading distributions for which the optimal rate allocation consists of a single codeword layer.
For fading distributions with continuous pdfs that are not quasiconcave, we conjecture that the expected-distortion--minimizing rate allocation remains discrete but may comprise multiple codeword layers.
By contrast,
a continuum of codeword layers is in general necessary
when maximizing expected capacity or minimizing expected distortion over a slowly fading channel
\cite{shamai03:bc_app_slow_fade_mimo, tian08:sr_bc_exp_dist_gaus, ng09:dist_min_glbc_sr}.
In a broader context, the techniques for source coding under fading side-information channels may be applied to improve quantize-and-forward schemes
\cite{katz09:coop_src_occa_relay} in wireless network transmissions, where the side information represents the auxiliary signals forwarded by a cooperating user as received via a fading channel.
In those cases where different distortion measures other than squared error are considered, however, different conclusions regarding the optimal number of source-coding layers may result.

\appendices

\section{Proof of Lemma~\ref{lem:VAR_XYW_XW_s}}
\label{sec:proof_VAR_XYW_XW_s}
\begin{IEEEproof}
The lemma follows from the minimum mean square error (MMSE) estimate of Gaussian random variables.
Let $X,\mathbf{W}$, where $\mathbf{W}\triangleq [W_1 \dots W_k]^T$, be distributed as
\begin{align}
\begin{bmatrix}\mathbf{W}\\X\end{bmatrix}
\sim \mathcal{N}
\biggl(\begin{bmatrix}\mathbf{\mu_W}\\\mu_X\end{bmatrix},
\begin{bmatrix}\mathbf{\Sigma_W}&\mathbf{\Sigma}_{\mathbf{W}X}\\
\mathbf{\Sigma}_{\mathbf{W}X}^T&\sigma_X^2\end{bmatrix}\biggr).
\end{align}
The conditional distribution is Gaussian, and the corresponding variance is
\begin{align}
\VAR[X|Y,\mathbf{W}] &= \sigma_X^2
-\begin{bmatrix}\mathbf{\Sigma}_{\mathbf{W}X} \\ \sqrt{s}\sigma_X^2\end{bmatrix}^T
\begin{bmatrix}\mathbf{\Sigma_W} & \sqrt{s}\mathbf{\Sigma}_{\mathbf{W}X} \\
\sqrt{s}\mathbf{\Sigma}_{\mathbf{W}X}^T & s\sigma_X^2+1\end{bmatrix}^{-1}
\begin{bmatrix}\mathbf{\Sigma}_{\mathbf{W}X} \\ \sqrt{s}\sigma_X^2\end{bmatrix}\\
&=\frac{\sigma_X^2-\mathbf{\Sigma}_{\mathbf{W}X}^T\mathbf{\Sigma_W}^{-1}\mathbf{\Sigma}_{\mathbf{W}X}}
{1+s(\mathbf{\Sigma}_{\mathbf{W}X}^T\mathbf{\Sigma_W}^{-1}\mathbf{\Sigma}_{\mathbf{W}X})}\\
&=\bigl(\VAR[X|\mathbf{W}]^{-1}+s\bigr)^{-1}.
\end{align}
\end{IEEEproof}

\section{Proof of Proposition~\ref{prop:cost_min_convex}}
\label{sec:proof_cost_min_convex}
\begin{IEEEproof}
Each of the inequality constraints in (\ref{eq:cvx_DV_log_V_Rx})--(\ref{eq:cvx_DV_Vi_Di}) is convex: i.e., it is of the form
\begin{align}
c_x(D_1,\dotsc,D_M,\, V_1,\dotsc,V_M) \leq c_c(D_1,\dotsc,D_M,\, V_1,\dotsc,V_M)
\end{align}
where $c_x(\cdot)$ is convex in $D_1,\dotsc,V_M$, and $c_c(\cdot)$ is concave in $D_1,\dotsc,V_M$.
In particular, in (\ref{eq:cvx_DV_Vi_si}), the right-hand side of each inequality constraint depends on only $V_{i-1}$.
Being twice-differentiable, its concavity can be verified by the second-order condition
\begin{align}
\label{eq:d2_dV2_concave}
\frac{d^2}{dV_{i-1}^2} (V_{i-1}^{-1}+s_i-s_{i-1})^{-1} = \frac{-2(s_i-s_{i-1})}{\bigl(1+(s_i-s_{i-1}) V_{i-1}\bigr)^3}
\end{align}
which is negative since $s_i > s_{i-1}$, $V_{i-1}>0$ as given in the problem formulation, for $i=2,\dotsc,M$.
Therefore, in (\ref{eq:cvx_DV_J})--(\ref{eq:cvx_DV_Vi_Di}), we minimize a convex function subject to a set of convex inequality constraints,
which is a convex optimization problem.
\end{IEEEproof}

\section{Convex Optimization Problem (\ref{eq:inf_disc_ED_pD})--(\ref{eq:inf_disc_ED_Di_si})}
\label{sec:inf_disc_exp_dist_min}

We show that (\ref{eq:inf_disc_ED_pD})--(\ref{eq:inf_disc_ED_Di_si}) is a convex problem.
For convenience, let us denote the left-hand side of (\ref{eq:inf_disc_ED_D_RX}) as follows:
\begin{align}
\Phi(\mathbf{D}) &\triangleq
-\frac{1}{2} \sum_{i=1}^\infty
\Bigl(
\log(D_{i-1}^{-1}+s_i-s_{i-1}) + \log D_i
\Bigr).
\end{align}
We show that $\Phi(\mathbf{D})$ is convex in $\mathbf{D}$.
Let $\mathbf{D}^{(1)}, \mathbf{D}^{(2)} \succ 0$ be two sets of element-wise positive distortion variables.
By the mean-value theorem, for some $\gamma\in[0,1]$, we have
\begin{align}
\label{eq:phi2_phi1_grad_hess}
\begin{split}
\Phi\bigl(\mathbf{D}^{(2)}\bigr) &= \Phi\bigl(\mathbf{D}^{(1)}\bigr)
+ \sum_{i=1}^\infty
\frac{\partial \Phi\bigl(\mathbf{D}^{(1)}\bigr)}{\partial D_i}
\bigl(D^{(2)}_i-D^{(1)}_i\bigr)\\
& \qquad
+ \sum_{i=1}^\infty \sum_{j=1}^\infty
\frac{\partial^2 \Phi\bigl(\mathbf{D}^{(1)} + \gamma(\mathbf{D}^{(2)}-\mathbf{D}^{(1)})\bigr)}
{\partial D_i \partial D_j}
\bigl(D^{(2)}_i-D^{(1)}_i\bigr) \bigl(D^{(2)}_j-D^{(1)}_j\bigr)
\end{split}\\
\label{eq:phi2_geq_phi1_grad}
&\geq \Phi\bigl(\mathbf{D}^{(1)}\bigr)
+ \sum_{i=1}^\infty
\frac{\partial \Phi\bigl(\mathbf{D}^{(1)}\bigr)}{\partial D_i}
\bigl(D^{(2)}_i-D^{(1)}_i\bigr)
\end{align}
which corresponds to the first-order convexity condition.
In (\ref{eq:phi2_geq_phi1_grad}), the inequality follows from:
\begin{align}
\frac{\partial^2 \Phi}{\partial D_i \partial D_j} &=
\begin{cases}
\frac{1}{2}\Bigl(\frac{s_{i+1}-s_i}{1+(s_{i+1}-s_i)D_i}\Bigr)^2, & i = j\\
0, & i \neq j
\end{cases}
\end{align}
which shows that the last term in (\ref{eq:phi2_phi1_grad_hess}) is nonnegative.
Each constraint in (\ref{eq:inf_disc_ED_Di_si}) is shown to be a convex set in (\ref{eq:d2_dV2_concave}).
Convexity is preserved under the intersection of a possibly infinite number of convex sets \cite[Section~2.3.1]{boyd04:convex_opt}.
The linear objective function in (\ref{eq:inf_disc_ED_pD}) is convex; therefore, (\ref{eq:inf_disc_ED_pD})--(\ref{eq:inf_disc_ED_Di_si}) is a convex optimization problem.

\section{KKT Optimality Conditions and Dual Function for (\ref{eq:inf_disc_ED_pD})--(\ref{eq:inf_disc_ED_Di_si})}
\label{sec:inf_disc_KKT_opt_dual_fn}

In the following, we characterize the KKT optimality conditions
\cite{luenberger69:opt_vect}
for the expected distortion minimization problem (\ref{eq:inf_disc_ED_pD})--(\ref{eq:inf_disc_ED_Di_si}).
First, we form the Lagrangian
\begin{align}
\label{eq:Lagrangian_disc}
L(\mathbf{D}, \boldsymbol\lambda, \mu) &= L_1(\mathbf{D}, \boldsymbol\lambda, \mu)
+ L_2(\mathbf{D}, \boldsymbol\lambda, \mu) + L_3(\mathbf{D}, \boldsymbol\lambda, \mu)\\
L_1(\mathbf{D}, \boldsymbol\lambda, \mu) &\triangleq \sum_{i=1}^\infty p_iD_i\\
L_2(\mathbf{D}, \boldsymbol\lambda, \mu) &\triangleq
\sum_{i=1}^\infty \lambda_i \Bigl(D_i - (D_{i-1}^{-1}+s_i-s_{i-1})^{-1}\Bigr)\\
L_3(\mathbf{D}, \boldsymbol\lambda, \mu) &\triangleq
\mu \Bigl(
-\frac{1}{2} \sum_{i=1}^\infty
\Bigl(
\log(D_{i-1}^{-1}+s_i-s_{i-1}) + \log D_i
\Bigr)
- R_X\Bigr)
\end{align}
where $\boldsymbol\lambda \triangleq [\lambda_1\; \lambda_2\; \dots]^T$,
and $\mu, \boldsymbol\lambda$ are the Lagrange multipliers, or dual variables, associated with inequalities (\ref{eq:inf_disc_ED_D_RX}), (\ref{eq:inf_disc_ED_Di_si}), respectively.
At optimality, the gradient of the Lagrangian vanishes:
\begin{align}
\label{eq:KKT_ED_dDi}
0 &= \frac{\partial L}{\partial D_i} = p_i + \lambda_i -\frac{\lambda_{i+1}}{\bigl(1 + (s_{i+1}-s_i)D_i\bigr)^2}
-\frac{\mu}{2}\frac{s_{i+1}-s_i}{1+(s_{i+1}-s_i)D_i},\quad i = 1,2,\dotsc
\end{align}
and the complementary slackness conditions hold:
\begin{align}
\label{eq:KKT_ED_lamb}
0 &= \lambda_i \Bigl(D_i - (D_{i-1}^{-1}+s_i-s_{i-1})^{-1}\Bigr),\quad i=1,2,\dots\\
\label{eq:KKT_ED_mu}
0 &=
\mu \Bigl(
-\frac{1}{2} \sum_{i=1}^\infty
\Bigl(
\log(D_{i-1}^{-1}+s_i-s_{i-1}) + \log D_i
\Bigr)
- R_X\Bigr).
\end{align}
The primal feasibility conditions are given by (\ref{eq:inf_disc_ED_D_RX}), (\ref{eq:inf_disc_ED_Di_si}),
and the dual feasibility conditions are
\begin{align}
\label{eq:KKT_ED_dual_feas}
\mu&\geq0,& \lambda_i&\geq0,\quad i=1,2,\dots.
\end{align}
Together,
(\ref{eq:inf_disc_ED_D_RX}), (\ref{eq:inf_disc_ED_Di_si}), and (\ref{eq:KKT_ED_dDi})--(\ref{eq:KKT_ED_dual_feas})
are the necessary and sufficient conditions for optimality in the convex problem
(\ref{eq:inf_disc_ED_pD})--(\ref{eq:inf_disc_ED_Di_si}).

The dual function of the expected distortion minimization problem is given by
\begin{align}
\label{eq:G_lambda_mu_L_D}
G(\boldsymbol\lambda,\mu) &= \inf_{\mathbf{D}} L(\mathbf{D},\boldsymbol\lambda,\mu).
\end{align}
Let $\hat{\mathbf{D}} \triangleq [\hat{D}_1\; \hat{D}_2\; \dots]^T$
denote the Lagrangian-minimizing $\mathbf{D}$ in (\ref{eq:G_lambda_mu_L_D}).
For $\boldsymbol\lambda\succeq0$, $\mu\geq0$,
$\hat{\mathbf{D}}$  can be determined from the KKT gradient conditions (\ref{eq:KKT_ED_dDi}):
\begin{align}
\label{eq:D_hat_i}
\begin{split}
\hat{D}_i(\boldsymbol\lambda,\mu) &= \frac{\mu(s_{i+1}-s_i) - 4(p_i+\lambda_i) + \sqrt{\mu^2(s_{i+1}-s_i)^2+16(p_i+\lambda_i)\lambda_{i+1}}}{4(s_{i+1}-s_i)(p_i+\lambda_i)},\\& \qquad i=1,2,\dotsc.
\end{split}
\end{align}
The dual function provides a lower bound to the solution of the optimization problem.
Let $J^*$ denote the minimum expected distortion in (\ref{eq:inf_disc_ED_pD})--(\ref{eq:inf_disc_ED_Di_si}).
For any $\tilde{\mathbf{D}}$ that is primal feasible by (\ref{eq:inf_disc_ED_D_RX}), (\ref{eq:inf_disc_ED_Di_si}),
and any dual feasible $\tilde{\boldsymbol\lambda} \succeq \mathbf{0}$, $\tilde{\mu}\geq0$, we have
\begin{align}
\label{eq:G_Jopt_J}
G(\tilde{\boldsymbol\lambda},\tilde{\mu}) \leq J^* \leq J(\tilde{\mathbf{D}}).
\end{align}
Furthermore, since the optimization problem is convex and Slater's condition holds, the duality gap is zero:
\begin{align}
G(\boldsymbol\lambda^*,\mu^*) = J^* = J(\mathbf{D}^*)
\end{align}
where $\mathbf{D}^*$ and $\boldsymbol\lambda^*,\mu^*$ respectively denote the optimal primal and dual variables
that satisfy the KKT optimality conditions
(\ref{eq:inf_disc_ED_D_RX}), (\ref{eq:inf_disc_ED_Di_si}),
and (\ref{eq:KKT_ED_dDi})--(\ref{eq:KKT_ED_dual_feas}).

\section{Proof of Proposition~\ref{prop:single_layer_lambda}}
\label{sec:proof_single_layer_lambda}

\begin{IEEEproof}
In the convex optimization problem
(\ref{eq:inf_disc_ED_pD})--(\ref{eq:inf_disc_ED_Di_si}),
the single-layer rate allocation $\mathcal{R}_{\bar{s}}$
is primal feasible by (\ref{eq:inf_disc_ED_D_RX}), (\ref{eq:inf_disc_ED_Di_si}).
Through construction of the dual variables $\bar\lambda_i$, $\bar\mu$ in
(\ref{eq:lambda_i_0})--(\ref{eq:mu_bar}),
the KKT gradient conditions (\ref{eq:KKT_ED_dDi})
and the complementary slackness conditions (\ref{eq:KKT_ED_lamb}), (\ref{eq:KKT_ED_mu}) are satisfied.
The dual feasibility condition $\bar\mu\geq0$ follows from the non-negativity of each term in the summation in (\ref{eq:mu_bar}).
Besides (\ref{eq:lambda_i_0}), dual feasibility of $\bar\lambda_i$ as given in (\ref{eq:prop_lambda_i_geq_0}) are the remaining necessary and sufficient conditions for optimality.
\end{IEEEproof}

\section{Proof of Proposition~\ref{prop:ct_single_layer_E_D}}
\label{sec:proof_ct_single_layer_E_D}

\begin{IEEEproof}
We assume the side-information channel has a fading distribution with a continuous pdf $f(s)$,
and we partition the continuum of fading state into subintervals
$[u_0,u_1],\, [u_1,u_2],\, \dotsc$ of lengths $\Delta u_1, \Delta u_2, \dotsc$,
with $\Delta u_i \triangleq u_i - u_{i-1}$, $u_0 = 0$.
Let us assume $\bar{s} > 0$; in the case of $\bar{s}=0$, it is interpreted as $\lim_{\bar{s}\rightarrow 0^+}$.
The partition boundaries are chosen such that $u_{\bar\imath}=\bar{s}$ for some index $\bar\imath$.
The fading realization falls within each subinterval with probability
$p_i = \int_{u_{i-1}}^{u_i} f(s) \;ds$, for $i=1,2,\dotsc$.
We consider two sets of discretized fading states.
In the first set, for each subinterval, we discretize the fading state to its worst-case realization:
$\{s_i\} = \mathbf{u}_\mathrm{w} \triangleq \{u_0, u_1,\dotsc\}$,
i.e., we have fading state $u_{i-1}$ with probability $p_i$.
In the second set, in a similar manner, for each subinterval we discretize the fading state to its best-case realization:
$\{s_i\} = \mathbf{u}_\mathrm{b} \triangleq \{u_1, u_2,\dotsc\}$,
where probability $p_i$ is associated with fading state $u_i$.

Next, we apply the single-layer rate allocation $\mathcal{R}_{\bar{s}}$
to the worst-case states $\mathbf{u}_\mathrm{w}$,
the continuous distribution $f(s)$,
and the best-case states $\mathbf{u}_\mathrm{b}$.
Since the decoder may arbitrarily add noise to the side-information channel, the expected distortions under $\mathcal{R}_{\bar{s}}$ are ordered as follows:
\begin{align}
\label{eq:J_b_J_c_J_w}
J_\mathrm{b}(\mathcal{R}_{\bar{s}}) \leq J_\mathrm{c}(\mathcal{R}_{\bar{s}}) \leq
J_\mathrm{w}(\mathcal{R}_{\bar{s}})
\end{align}
where $J_\mathrm{b}(\mathcal{R}_{\bar{s}})$, $J_\mathrm{c}(\mathcal{R}_{\bar{s}})$, $J_\mathrm{w}(\mathcal{R}_{\bar{s}})$ represent
the expected distortions under $\mathcal{R}_{\bar{s}}$ for the best-case states, continuous distribution, and worst-case states, respectively.

Finally, consider the inequalities in (\ref{eq:J_b_J_c_J_w}) in the limit of small subintervals
as $\norm{\mathbf{u}} \rightarrow 0$,
where $\norm{\mathbf{u}} \triangleq \max \Delta u_i$.
As $f(s)$ is continuous, by the mean-value theorem, we have
\begin{align}
\label{eq:p_i_f_mean_value}
p_i &= \int_{u_{i-1}}^{u_i} f(s) \;ds = (u_i-u_{i-1})f(u_i^*)
\end{align}
for some $u_i^* \in [u_{i-1}, u_i]$ that achieves the mean value of $f(\cdot)$ within the partition $[u_{i-1}, u_i]$.
Apply (\ref{eq:p_i_f_mean_value}) to (\ref{eq:D_bar_i}), (\ref{eq:E_D_bar}), we have
\begin{align}
\label{eq:Jb_sum_inv_delta_u}
\lim_{\norm{\mathbf{u}} \rightarrow 0} J_\mathrm{b}(\mathcal{R}_{\bar{s}})
&= \lim_{\norm{\mathbf{u}} \rightarrow 0}\:
\sum_{i=1}^{\bar\imath-1} \frac{(u_i-u_{i-1})f(u_i^*)}{\sigma_X^{-2} + u_i}
+ \sum_{i=\bar\imath}^\infty \frac{(u_i-u_{i-1})f(u_i^*)}{(\sigma_X^{-2}+\bar{s})e^{2R_X} + u_i - \bar{s}}\\
\label{eq:Jb_Jw_int_f_u}
&= \int_0^{\bar{s}} \frac{f(u)}{\sigma_X^{-2}+u}\;du
+ \int_{\bar{s}}^\infty \frac{f(u)}{(\sigma_X^{-2}+\bar{s})e^{2R_X}+u-\bar{s}} \;du\\
\label{eq:sum_ibar1_inv_delta_u}
&= \lim_{\norm{\mathbf{u}} \rightarrow 0}\:
\sum_{i=1}^{\bar\imath-1} \frac{(u_i-u_{i-1})f(u_i^*)}{\sigma_X^{-2} + u_{i-1}}
+ \sum_{i=\bar\imath}^\infty \frac{(u_i-u_{i-1})f(u_i^*)}{(\sigma_X^{-2}+\bar{s})e^{2R_X} + u_{i-1} - \bar{s}}\\
\label{eq:lim_Delta_u_Jw_Rs}
&= \lim_{\norm{\mathbf{u}} \rightarrow 0} J_\mathrm{w}(\mathcal{R}_{\bar{s}}).
\end{align}
Since from (\ref{eq:Jb_sum_inv_delta_u})--(\ref{eq:lim_Delta_u_Jw_Rs}) the limits of
$J_\mathrm{b}(\mathcal{R}_{\bar{s}})$, $J_\mathrm{w}(\mathcal{R}_{\bar{s}})$ coincide and equal
(\ref{eq:Jb_Jw_int_f_u}),
from (\ref{eq:J_b_J_c_J_w}) $J_\mathrm{c}(\mathcal{R}_{\bar{s}})$ also equals this limit,
which is given by (\ref{eq:E_D_bar_f_s}) in the main body of the text.
\end{IEEEproof}

\section{Proof of Proposition~\ref{prop:ct_qconcave_single_opt}}
\label{sec:proof_ct_qconcave_single_opt}

\begin{IEEEproof}
Consider the set of best-case and worst-case discretized fading states described in Appendix~\ref{sec:proof_ct_single_layer_E_D}.
We have the following inequalities:
\begin{align}
\label{eq:Gb_Jb_Jc_Jw_JwDw}
G_\mathrm{b}\bigl(\bar\lambda(\mathbf{u}_\mathrm{b}),\bar\mu\bigr)
\stackrel{(a)}{\leq} J_\mathrm{b}^*
\stackrel{(b)}{\leq} J_\mathrm{c}^*
\stackrel{(c)}{\leq} J_\mathrm{w}^*
\stackrel{(d)}{\leq} J_\mathrm{w}(\mathcal{R}_{\bar{s}})
\end{align}
where $J_\mathrm{b}^*$, $J_\mathrm{c}^*$, $J_\mathrm{w}^*$ denote the minimum expected distortion under the best-case states, continuous distribution, and worst-case states, respectively;
$G_\mathrm{b}(\cdot)$
is the dual function under the best-case states;
$\bar\lambda(\mathbf{u}_\mathrm{b})$ denotes applying $\bar\lambda(\cdot)$ on $\mathbf{u}_\mathrm{b}$
element-wise;
and $J_\mathrm{w}(\mathcal{R}_{\bar{s}})$ is as defined in (\ref{eq:J_b_J_c_J_w}).
In (\ref{eq:Gb_Jb_Jc_Jw_JwDw}), $(b),(c)$ follow from the ordering on the sets of the fading states,
and $(a),(d)$ follow from the duality bounds (\ref{eq:G_Jopt_J}):
$\bar\lambda(\mathbf{u}_\mathrm{b})$ is dual feasible by the assumption $\bar\lambda(s)\geq0$;
$\bar\mu$ is dual feasible since the integrand in (\ref{eq:mu_bar_f_s}) is nonnegative;
and $\mathcal{R}_{\bar{s}}$, being a valid rate allocation, induces a set of distortion variables that is primal feasible.

Next, we consider $G_\mathrm{b}\bigl(\bar\lambda(\mathbf{u}_\mathrm{b}),\bar\mu\bigr)$
in the limit of $\norm{\mathbf{u}} \rightarrow 0$.
Writing (\ref{eq:G_lambda_mu_L_D}) in terms of (\ref{eq:Lagrangian_disc}), recall that
\begin{align}
\label{eq:Gb_L1_L2_L3}
\begin{split}
G_\mathrm{b}\bigl(\bar\lambda(\mathbf{u}_\mathrm{b}),\bar\mu\bigr)
&= L_1\bigl(\hat{\mathbf{D}}\bigl(\bar\lambda(\mathbf{u}_\mathrm{b}),\bar\mu\bigr),\bar\lambda(\mathbf{u}_\mathrm{b}),\bar\mu\bigr)
+ L_2\bigl(\hat{\mathbf{D}}\bigl(\bar\lambda(\mathbf{u}_\mathrm{b}),\bar\mu\bigr),\bar\lambda(\mathbf{u}_\mathrm{b}),\bar\mu\bigr)\\
&\quad + L_3\bigl(\hat{\mathbf{D}}\bigl(\bar\lambda(\mathbf{u}_\mathrm{b}),\bar\mu\bigr),\bar\lambda(\mathbf{u}_\mathrm{b}),\bar\mu\bigr)
\end{split}
\end{align}
where $\hat{\mathbf{D}}\bigl(\bar\lambda(\mathbf{u}_\mathrm{b}),\bar\mu\bigr)$ is as given in (\ref{eq:D_hat_i}).
Note that each $\hat{D}_i\bigl(\bar\lambda(\mathbf{u}_\mathrm{b}),\bar\mu\bigr)$ depends on $\bar{\lambda}(u_i)$ and $\bar{\lambda}(u_{i+1})$.
Since $\bar{\lambda}(s)$ is continuous and differentiable over $s>0$,
its values at the partition boundaries are related through the mean-value theorem:
\begin{align}
\label{eq:lamba_bar_mean_value}
\bar{\lambda}(u_{i+1}) &= \bar{\lambda}(u_i) + (u_{i+1}-u_i) \bar{\lambda}'(u_i^\dagger)
\end{align}
for some $u_i^\dagger \in [u_i, u_{i+1}]$.
Consider the first term in (\ref{eq:Gb_L1_L2_L3}):
\begin{align}
\label{eq:L_1_p_i_D_hat_i}
\lim_{\norm{\mathbf{u}} \rightarrow 0}
L_1\bigl(\hat{\mathbf{D}}\bigl(\bar\lambda(\mathbf{u}_\mathrm{b}), \bar\mu\bigr),
\bar\lambda(\mathbf{u}_\mathrm{b}), \bar\mu\bigr)
&= \lim_{\norm{\mathbf{u}} \rightarrow 0}\: \sum_{i=1}^\infty p_i \hat{D}_i\bigl(\bar\lambda(\mathbf{u}_\mathrm{b}), \bar\mu\bigr)\\
\label{eq:int_f_lambda_bar_f_du}
&= \int_0^\infty f(u) \frac{\bar\lambda'(u)-f(u)+\bar\mu/2}{2\bar\lambda(u)} \;du\\
\label{eq:int_f_u_exp_dist}
&= \int_0^{\bar{s}} \frac{f(u)}{\sigma_X^{-2}+u}\;du
+ \int_{\bar{s}}^\infty \frac{f(u)}{(\sigma_X^{-2}+\bar{s})e^{2R_X}+u-\bar{s}} \;du\\
\label{eq:L1_Jw}
&= \lim_{\norm{\mathbf{u}} \rightarrow 0} J_\mathrm{w}(\mathcal{R}_{\bar{s}})
\end{align}
where (\ref{eq:int_f_lambda_bar_f_du}) follows from applying
(\ref{eq:p_i_f_mean_value}), (\ref{eq:lamba_bar_mean_value}) to (\ref{eq:L_1_p_i_D_hat_i});
and (\ref{eq:L1_Jw}) follows from (\ref{eq:Jb_Jw_int_f_u}).
In (\ref{eq:int_f_lambda_bar_f_du}),
$\bar{\lambda}(\cdot)$ is as defined in (\ref{eq:lambda_bar_s_1}), (\ref{eq:lambda_bar_s_2}),
and its derivative is given as follows:
\begin{align}
\label{eq:d_lambda_s_1}
\bar\lambda'(s) &= \frac{\bar\mu}{2}\frac{\bar{s}-2s-\sigma_X^{-2}}{\sigma_X^{-2}+\bar{s}}+f(s)
-\int_s^{\bar{s}} 2f(t)\frac{\sigma_X^{-2}+s}{(\sigma_X^{-2}+t)^2} \;dt,
\qquad 0\leq s < \bar{s}\\
\label{eq:d_lambda_s_2}
\begin{split}
\bar\lambda'(s) &= -\frac{\bar\mu}{2}-\frac{\bar\mu(s-\bar{s})}{(\sigma_X^{-2}+\bar{s})e^{2R_X}}+f(s)\\
&\qquad +\int_{\bar{s}}^s 2f(t) \frac{(\sigma_X^{-2}+\bar{s})e^{2R_X}+s-\bar{s}}{\bigl((\sigma_X^{-2}+\bar{s})e^{2R_X}+t-\bar{s}\bigr)^2}\;dt,
\qquad s\geq\bar{s}.
\end{split}
\end{align}
Substituting (\ref{eq:lambda_bar_s_1}), (\ref{eq:lambda_bar_s_2}), (\ref{eq:d_lambda_s_1}), (\ref{eq:d_lambda_s_2}) into (\ref{eq:int_f_lambda_bar_f_du}),
note that the factor in the integrand (defined as $D(s)$ below)
simplifies to the following expressions which are independent of $f(s)$:
\begin{align}
D(s) &\triangleq
\frac{\bar\lambda'(s)-f(s)+\bar\mu/2}{2\bar\lambda(s)}\\
&= \begin{cases}
(\sigma_X^{-2} + s)^{-1}, & 0\leq s < \bar{s}\\
\big((\sigma_X^{-2}+\bar{s})e^{2R_X} + s - \bar{s}\bigr)^{-1},
& s \geq \bar{s}.
\end{cases}
\end{align}

For the second term in (\ref{eq:Gb_L1_L2_L3}),
note that $\bar\lambda(u_{\bar\imath})=0$,
and hence $i=\bar\imath$ can be excluded from the following summation:
\begin{align}
\label{eq:sum_lambda_D_d_delta}
\begin{split}
&\lim_{\norm{\mathbf{u}} \rightarrow0}
L_2\bigl(\hat{\mathbf{D}}\bigl(\bar\lambda(\mathbf{u}_\mathrm{b}), \bar\mu\bigr),
\bar\lambda(\mathbf{u}_\mathrm{b}), \bar\mu\bigr)\\
&\quad = \lim_{\norm{\mathbf{u}} \rightarrow0}\: \sum_{i\neq\bar\imath} \bar\lambda(u_i)
\Bigl(\hat{D}_i\bigl(\bar\lambda(\mathbf{u}_\mathrm{b}), \bar\mu\bigr)
- \Bigl(\hat{D}_{i-1}\bigl(\bar\lambda(\mathbf{u}_\mathrm{b}), \bar\mu\bigr)^{-1}
+ u_i - u_{i-1}
\Bigr)^{-1}\Bigr)
\end{split}\\
\label{eq:int_lambda_D_d_D2}
&\quad =  \int_{\backslash\bar{s}} \bar\lambda(u)\bigl(D'(u)+D(u)^2\bigr) \;du\\
\label{eq:L2_0}
&\quad = 0
\end{align}
where
$\hat{D}_0(\cdot) \triangleq \sigma_X^2$, and
(\ref{eq:p_i_f_mean_value}), (\ref{eq:lamba_bar_mean_value}) are applied
in (\ref{eq:sum_lambda_D_d_delta});
the notation $\int_{\backslash\bar{s}} \cdot \;du$ in (\ref{eq:int_lambda_D_d_D2})
is defined as
\begin{align}
\label{eq:int_backslash_sbar_def}
\int_{\backslash\bar{s}} g(u) \;du \triangleq \int_0^{\bar{s}^-} g(u) \;du + \int_{\bar{s}^+}^\infty g(u) \;du
\end{align}
convergence of the partition summation follows from $f(s)$, $\bar\lambda(s)$, $\bar\lambda'(s)$ being continuous
when the mean-value theorem is applied to $\hat{D}_i(\cdot)$ in (\ref{eq:D_hat_i});
and (\ref{eq:L2_0}) follows from
$D'(s)+D(s)^2=0$
over the intervals $0<s<\bar{s}$ and $s>\bar{s}$.
Finally, consider the third term in (\ref{eq:Gb_L1_L2_L3}):
\begin{align}
\label{eq:L3_log_D1_Di_RX}
\begin{split}
&\lim_{\norm{\mathbf{u}} \rightarrow0}
L_3\bigl(\hat{\mathbf{D}}\bigl(\bar\lambda(\mathbf{u}_\mathrm{b}), \bar\mu\bigr),
\bar\lambda(\mathbf{u}_\mathrm{b}), \bar\mu\bigr)\\
&\quad = \lim_{\norm{\mathbf{u}} \rightarrow0}\:
\bar{\mu} \Biggl(
-\frac{1}{2} \sum_{i=1}^\infty
\biggl(
\log\Bigl(\hat{D}_{i-1}\bigl(\bar\lambda(\mathbf{u}_\mathrm{b}), \bar\mu\bigr)^{-1}+u_i-u_{i-1}\Bigr)
+ \log \hat{D}_i\bigl(\bar\lambda(\mathbf{u}_\mathrm{b}), \bar\mu\bigr)
\biggr)
- R_X\Biggr)
\end{split}\\
\label{eq:u_int_D_dD_u_RX}
&\quad = \frac{\bar\mu}{2}\int_{\backslash\bar{s}} D(u)+\frac{D'(u)}{D(u)} \;du
+\bar\mu(R_X-R_X)\\
\label{eq:L3_0}
&\quad = 0
\end{align}
where
(\ref{eq:p_i_f_mean_value}), (\ref{eq:lamba_bar_mean_value}) are applied in (\ref{eq:L3_log_D1_Di_RX});
the last term in (\ref{eq:u_int_D_dD_u_RX}) follows from noting that
$\hat{D}_{\bar{\imath}-1}(\cdot) \rightarrow (\sigma_X^{-2} + \bar{s})^{-1}$ and
$\hat{D}_{\bar{\imath}}(\cdot) \rightarrow \big((\sigma_X^{-2}+\bar{s})e^{2R_X}\bigr)^{-1}$;
the notation $\int_{\backslash\bar{s}} \cdot \;du$ is as given in (\ref{eq:int_backslash_sbar_def}),
and its convergence follows from continuity of $f(s)$, $\bar\lambda(s)$, $\bar\lambda'(s)$;
(\ref{eq:L3_0}) follows from $D(s)+D'(s)/D(s)=0$ over the intervals $0<s<\bar{s}$ and $s>\bar{s}$.
Combining (\ref{eq:Gb_L1_L2_L3}), (\ref{eq:L1_Jw}), (\ref{eq:L2_0}), (\ref{eq:L3_0}), we have:
$\lim_{\Delta u\rightarrow0} G_\mathrm{b}\bigl(\bar\lambda(\mathbf{u}_\mathrm{b}),\bar\mu\bigr)
= \lim_{\Delta u \rightarrow 0} J_\mathrm{w}(\mathcal{R}_{\bar{s}})$.
Since the inequalities in (\ref{eq:Gb_Jb_Jc_Jw_JwDw}) coincide in the limit,
the minimum expected distortion $J_\mathrm{c}^*$ is given by (\ref{eq:int_f_u_exp_dist}),
which is achievable by $\mathcal{R}_{\bar{s}}$ by Proposition~\ref{prop:ct_single_layer_E_D}.
\end{IEEEproof}

\section{Proof of Proposition~\ref{prop:f_s_qc_single_layer}}
\label{sec:proof_f_s_qc_single_layer}

\begin{IEEEproof}
We show that the single-layer rate allocation satisfies the optimality conditions (\ref{eq:lambda_bar_s_geq_0}) in Proposition~\ref{prop:ct_qconcave_single_opt}.
It will be useful to compare $f(s)$ against $\bar\mu/2$,
so we bring $\bar\mu/2$ inside in the integrals in (\ref{eq:lambda_bar_s_1}), (\ref{eq:lambda_bar_s_2}), and rewrite the expressions as
\begin{align}
\label{eq:labmda_s_w_n}
\frac{\bar\lambda(s)}{w_n(s)} &=
\int_{\bar{s}}^s \frac{f(t)-\bar\mu/2}{w_n(t)}\;dt
\end{align}
where $n=1$ for $0\leq s < \bar{s}$, $n=2$ for $s\geq\bar{s}$,
with $w_n(s)\geq0$ given as follows:
\begin{align}
\label{eq_w_1_s}
w_1(s) &= (\sigma_X^{-2}+s)^2\\
\label{eq_w_2_s}
w_2(s) &= \bigl((\sigma_X^{-2}+\bar{s})e^{2R_X}+s-\bar{s}\bigr)^2.
\end{align}
We now show the non-negativity of (\ref{eq:labmda_s_w_n}).
Solving for $\alpha^*$ in (\ref{eq:alpha_f_s_a}) and comparing its solution against the right-hand side of (\ref{eq:mu_bar_f_s}), we recognize that $\alpha^*=\bar\mu/2$.
Thus for $0\leq s<\bar{s}^*$, being outside of the superlevel set $[s_\mathrm{a},s_\mathrm{b}]$,
we have $f(s)<\bar\mu/2$ in the integrand in (\ref{eq:labmda_s_w_n}):
hence $\bar\lambda(s)>0$.
For $s\geq\bar{s}^*$ but $s\leq s_\mathrm{b}$, we have $f(s)\geq\bar\mu/2$: hence $\bar\lambda(s)\geq0$.
Finally, for $s>s_\mathrm{b}$, note that $\bar\lambda(s)/w_2(s)$ is monotonically decreasing, but it never descends below zero:
$\lim_{s\rightarrow\infty} \bar\lambda(s)/w_2(s) = 0$,
which is a consequence of how $\alpha^*$ is constructed as specified in (\ref{eq:alpha_f_s_a}).
\end{IEEEproof}

\section*{Acknowledgment}

The authors thank Erik Ordentlich for providing detailed feedback on the manuscript and his valuable technical suggestions during the review process.



\begin{IEEEbiographynophoto}{Chris~T.~K.~Ng}
(S'99--M'07) received the B.A.Sc.\ degree in engineering science from the University of Toronto, Toronto, ON, Canada.
He received the M.S.\ and Ph.D.\ degrees in electrical engineering from Stanford University, Stanford, CA.
Dr.~Ng was a Member of Technical Staff at Bell Labs, Alcatel-Lucent, in Holmdel, NJ.
From 2007 to 2008, he was a Postdoctoral Researcher in the Department of Electrical Engineering and Computer Science at the Massachusetts Institute of Technology,
Cambridge, MA.
His research interests include cooperative communications, joint source-channel coding, cross-layer wireless network design, optimization, and network information theory.
Dr.~Ng was a recipient of the 2007 IEEE International Symposium on Information
Theory Best Student Paper Award, and a recipient of a Croucher Foundation
Fellowship in 2007.
\end{IEEEbiographynophoto}

\begin{IEEEbiographynophoto}{Chao~Tian}
(S'00--M'05) received the B.E.\ degree in Electronic Engineering
from Tsinghua University, Beijing, China, in 2000 and the M.S.\ and Ph.D.\
degrees in Electrical and Computer Engineering from Cornell University,
Ithaca, NY in 2003 and 2005, respectively.

Dr.~Tian was a postdoctoral researcher at Ecole Polytechnique Federale de
Lausanne (EPFL) from 2005 to 2007. He joined AT\&T Labs--Research,
Florham Park, New Jersey in 2007, where he is now a Senior Member of
Technical Staff. His research interests include multi-user information theory,
joint source-channel coding, quantization design and analysis, as well as
image/video coding and processing. Dr.~Tian is currently an associated editor
for IEEE Signal Processing Letters.
\end{IEEEbiographynophoto}

\begin{IEEEbiographynophoto}{Andrea~Goldsmith}
is a professor of Electrical Engineering at Stanford University, and was previously an assistant professor of Electrical Engineering at Caltech. She co-founded Accelera Mobile Broadband, Inc.\ and Quantenna Communications Inc., and has previously held industry positions at Maxim Technologies, Memorylink Corporation, and AT\&T Bell Laboratories. Dr.~Goldsmith is a Fellow of the IEEE and of Stanford, and she has received several awards for her work, including the IEEE Communications Society and Information Theory Society joint paper award, the National Academy of Engineering Gilbreth Lecture Award, the IEEE Wireless Communications Technical Committee Recognition Award, the Alfred P.\ Sloan Fellowship, and the Silicon Valley/San Jose Business Journal's Women of Influence Award. Her research includes work on wireless information and communication theory, multihop wireless networks, cognitive radios, sensor networks, distributed control systems, ``green'' wireless system design, and applications of communications and signal processing to biology and neuroscience. She is author of the book ``Wireless Communications'' and co-author of the books ``MIMO Wireless Communications'' and ``Principles of Cognitive Radio,'' all published by Cambridge University Press. She received the B.S., M.S.\ and Ph.D.\ degrees in Electrical Engineering from U.C.\ Berkeley.

Dr.~Goldsmith has served as associate editor for the IEEE Transactions on Information Theory and as editor for the Journal on Foundations and Trends in Communications and Information Theory and in Networks. She previously served as an editor for the IEEE Transactions on Communications and for the IEEE Wireless Communications Magazine, as well as guest editor for several IEEE journal and magazine special issues.
Dr.~Goldsmith participates actively in committees and conference organization for the IEEE Information Theory and Communications Societies and has served on the Board of Governors for both societies. She is a Distinguished Lecturer for both societies, served as the President of the IEEE Information Theory Society in 2009, founded and chaired the student committee of the IEEE Information Theory society, and currently chairs the Emerging Technology Committee and is a member of the Strategic Planning Committee in the IEEE Communications Society. At Stanford she received the inaugural University Postdoc Mentoring Award, served as Chair of its Faculty Senate, and currently serves on its Faculty Senate and on its Budget Group.
\end{IEEEbiographynophoto}

\begin{IEEEbiographynophoto}{Shlomo~Shamai~(Shitz)}
received the B.Sc., M.Sc., and Ph.D.\ degrees in
electrical engineering from the Technion---Israel Institute of Technology,
in 1975, 1981 and 1986 respectively.

During 1975--1985 he was with the Communications Research Labs,
in the capacity of a Senior Research Engineer. Since 1986 he is with
the Department of Electrical Engineering, Technion---Israel Institute of
Technology, where he is now a Technion Distinguished Professor,
and holds the William Fondiller Chair of Telecommunications.
His research interests encompasses a wide spectrum of topics in information
theory and statistical communications.

Dr.~Shamai (Shitz) is an IEEE Fellow, and the recipient of the 2011
Claude E.\ Shannon Award.
He is the recipient of the 1999 van der Pol Gold Medal of the Union Radio
Scientifique Internationale (URSI), and a co-recipient of the 2000 IEEE
Donald G.\ Fink Prize Paper Award, the 2003, and
the 2004 joint IT/COM societies paper award, the 2007 IEEE Information
Theory Society Paper Award, the 2009 European Commission FP7, Network of
Excellence in Wireless COMmunications (NEWCOM++) Best Paper Award,
and the 2010 Thomson Reuters Award for International Excellence
in Scientific Research.  He is also the recipient of
1985 Alon Grant for distinguished young scientists and the 2000 Technion Henry
Taub Prize for Excellence in Research.
He has served as Associate Editor for the Shannon Theory of the IEEE
Transactions on Information Theory, and has also served twice on the
Board of Governors of the Information Theory Society.
He is a member of the Executive Editorial Board of the IEEE Transactions
on Information Theory.
\end{IEEEbiographynophoto}

\end{document}